\documentclass[12pt]{iopart}

\usepackage{graphicx}
\usepackage{hyperref}
\usepackage{times}
\usepackage{xcolor}
\usepackage{microtype}
\usepackage{subfigure}
\usepackage{booktabs}
\usepackage{amsthm}
\usepackage{bm}
\usepackage[ruled,vlined]{algorithm2e}
\usepackage{algorithmic}
\usepackage{orcidlink}
\usepackage{multicol}

\usepackage{iopams}  

\begin{document}

\title[DINAMO: Dynamic and INterpretable Anomaly MOnitoring]{DINAMO: Dynamic and INterpretable Anomaly MOnitoring for Large-Scale Particle Physics Experiments}

\author{%
  Arsenii~Gavrikov$^{\orcidlink{0000-0002-6741-5409}}$\footnotemark[1] \\ \address{INFN, Sezione di Padova e Università di Padova, Dipartimento di Fisica e Astronomia, Italy} \ead{agavriko@cern.ch} 
}

\author{%
  Julián~García~Pardiñas$^{\orcidlink{0000-0003-2316-8829}}$\footnotemark[1] \\ \address{Department of Experimental Physics, European Organization for Nuclear Research (CERN), 1211 Geneva 23, Switzerland, now at Laboratory for Nuclear Science, Massachusetts Institute of Technology (MIT), 77 Massachusetts Ave, Cambridge, MA 02139, USA} \ead{julian.garcia.pardinas@cern.ch}
}

\author{%
  Alberto~Garfagnini$^{\orcidlink{0000-0003-0658-1830}}$ \\ \address{INFN, Sezione di Padova e Università di Padova, Dipartimento di Fisica e Astronomia, Italy} \ead{alberto.garfagnini@cern.ch}
}

\footnotetext[1]{Authors to whom any correspondence should be addressed.}

%\vspace{10pt}
%\begin{indented}
%\item[]August 2017 (minor update March 2024)
%\end{indented}

\begin{abstract}
Ensuring reliable data collection in large-scale particle physics experiments demands Data Quality Monitoring (DQM) procedures to detect possible detector malfunctions and preserve data integrity. Traditionally, this resource-intensive task has been handled by human shifters who struggle with frequent changes in operational conditions. We present DINAMO: a novel, interpretable, robust, and scalable DQM framework designed to automate anomaly detection in time-dependent settings. Our approach constructs evolving histogram templates with built-in uncertainties, featuring both a statistical variant --- extending the classical Exponentially Weighted Moving Average (EWMA) --- and a machine learning (ML)-enhanced version that leverages a transformer encoder for improved adaptability. Experimental validations on synthetic datasets demonstrate the high accuracy, adaptability, and interpretability of these methods. The statistical variant is being commissioned in the LHCb experiment at the Large Hadron Collider, underscoring its real-world impact. The code used in this study is available at~\url{https://github.com/ArseniiGav/DINAMO}.
\end{abstract}


%\maketitle

\section{Introduction}
\label{sec:introduction}

The integrity of experimental data is fundamental to the success of any scientific discovery in physics. In large-scale particle physics experiments, particularly at the Large Hadron Collider (LHC)~\cite{ATLAS:2008xda,CMS:2008xjf,LHCb:2008vvz,ALICE:2008ngc}, vast amounts of data are collected continuously, requiring robust validation mechanisms to ensure their quality. The complex, multicomponent nature of such experimental instruments makes various types of failures at different levels of data collection inevitable. To ensure that the physics experiments bring reliable scientific discoveries, strict quality criteria are applied to all collected data, with only validated data proceeding to subsequent physics analyses. This monitoring and data certification process is done through data quality monitoring (DQM) systems~\cite{ai_for_hep_chapter5}.
Traditionally, DQM systems rely on dedicated human shifters who review aggregated information from specific detector components or variables to identify anomalous behavior in detector outputs. Following specific predetermined guidelines, the identification is then performed by comparing collected data to the expected (so-called reference) detector outputs. The references are typically constructed manually by human experts based on previously validated good data. This manual approach presents several challenges:

\begin{itemize}
    \item Firstly, it requires significant human resources. The complexity and volume of data often exceed what human operators can efficiently process leading to potential errors and inconsistencies. Moreover, interpretation differences between shifters can result in inconsistent and untraceable quality assessments.

    \item Secondly, while human shifters are often not detector experts, reference construction and guidelines formulation require consistent attention from a small group of detector experts. This is particularly problematic during dynamic operational periods, such as the commissioning of a new detector component or a major software update.
\end{itemize}

DQM typically operates on two levels: an \emph{online} stage, which runs in near real-time during data collection to quickly flag issues~\cite{online_dqm_lhcb, online_dqm_atlas}, and an \emph{offline} stage, which offers a more thorough evaluation after a data-taking run has concluded~\cite{offline_dqm_atlas}. Both stages are similar but differ in a few important aspects: the online stage requires immediate shifter response and operates under computational constraints that limit the scope of monitored features, whereas the offline stage allows comprehensive analysis without the direct time pressure. Usually, the final certification of the collected data is done at the offline stage. At both levels, an effective automated DQM solution must satisfy several key requirements:

\begin{itemize}
\item \textbf{High Accuracy:} The system must reliably distinguish between \emph{good} and \emph{bad} data. False positives waste valuable physics data, while false negatives risk compromising scientific conclusions.

\item \textbf{Targeted Human Oversight:} While human expertise remains essential for complex diagnostics, it should be engaged selectively. Data statistically consistent with nominal references can be automatically approved, freeing human resources for more critical tasks.

\item \textbf{Adaptability:} References and their uncertainties should ideally update seamlessly as operational conditions evolve, without requiring manual intervention.

\item \textbf{Interpretability:} The system's decisions must be transparent and verifiable by detector experts. Algorithms should provide valuable information to help identify anomaly causes and enable verification of decision-making processes.
\end{itemize}

Data quality monitoring intersects with broader anomaly detection research. In particle physics, traditional approaches rely on reference-based statistical tests (e.g., $\chi^2$, Kolmogorov--Smirnov) against pre-certified ``golden'' datasets. More recently, machine learning techniques have gained prominence, including supervised CNN classifiers, autoencoders, and LSTMs~\cite{ai_for_hep_chapter5}. The CMS Collaboration~\cite{CMS:2008xjf}, for instance, has employed autoencoders to identify anomalous calorimeter readouts~\cite{CMSECAL:2023fvz}, while a novel approach called ``AutoDQM''~\cite{Brinkerhoff:2025rob} combines statistical techniques with unsupervised machine learning. 

However, existing automated approaches typically assume static operational regimes, with ML-based methods often requiring re-training when conditions change. While effective for relatively stable conditions, these methods become less pragmatic during detector commissioning periods characterized by frequent and unpredictable operational changes. This limitation is particularly evident during commissioning periods following major upgrades, such as those underway for the High-Luminosity LHC era~\cite{Apollinari:2017lan}. These periods are characterized by frequent, sometimes unexpected, modifications to detector configurations. While reinforcement learning methods have been proposed to handle time-dependency~\cite{Parra:2024nue} and further automate the DQM process, their complexity and the substantial time required for experiments can be barriers to adoption.

Beyond particle physics, many time series anomaly-detection frameworks track evolving data distributions via classical methods like the Exponentially Weighted Moving Average (EWMA) \cite{264b1310-db27-36ce-b68f-3e439b03c2b5}, which can flag deviations from a running mean but may struggle with nonlinear drifts or high-dimensional data. To model more complex behaviors, machine-learning architectures --- ranging from RNNs and LSTMs~\cite{10.1145/3219819.3219845,10.1145/3292500.3330672} to Transformers~\cite{NIPS2017_3f5ee243,Xu2021AnomalyTT} --- have become prominent, although they often require frequent re-training or threshold tuning when distributions shift. Online or incremental learning~\cite{Widmer1996} offers an option for parameters to adapt continuously to new data in dynamic settings. Finally, a persistent challenge lies in the interpretability of the tools, as many contemporary algorithms produce opaque anomaly scores~\cite{10.1145/3691338}. In safety-critical domains like experimental particle physics, explainable anomaly detection~\cite{8466590} --- with explicit uncertainties or feature-level deviation maps—remains essential for building trust and enabling root-cause analysis.

In this paper, we introduce \textbf{DINAMO} (\underline{D}ynamic and \underline{IN}terpretable \underline{A}nomaly \underline{MO}nitoring), a framework comprising two complementary approaches for automating DQM in particle physics. Our solution emphasizes adaptability to evolving detector conditions and interpretability while maintaining robustness and technical simplicity. The framework includes a statistical method based on a modification of EWMA (DINAMO-S) and a machine learning (ML) method that substitutes the EWMA with a transformer encoder (DINAMO-ML). Although developed with the LHCb experiment~\cite{LHCb:2008vvz} as the primary use case, DINAMO is designed to be general enough for deployment across different experiments in both online and offline regimes. Demonstrating its practical impact in operational settings, DINAMO-S is currently being commissioned at the LHCb experiment for offline DQM. To enable comprehensive evaluation of the methods, we developed a synthetic data generator that reproduces gradual drifts, sudden operational shifts, systematic uncertainties, and other features of real-world detector data.

The rest of this paper is organized as follows:~\autoref{sec:methods} presents details of the DINAMO approaches, their technical aspects and mathematical formalization.~\autoref{sec:goals_metrics} presents the evaluation metrics.~\autoref{sec:data} describes the synthetic data generator and its features.~\autoref{sec:results} is dedicated to the results and to the methods comparison.~\autoref{sec:summary} presents the summary, discusses the limitations of the methods and the practical implementation at LHCb.

\section{Methods}
\label{sec:methods}

In this section, we describe the proposed DINAMO-S and DINAMO-ML algorithms. DINAMO learns statistical representations of nominal data in histogram format, maintaining templates with explicit uncertainties that distinguish between statistical fluctuations and genuine anomalies. These templates serve as references for newly observed histograms, triggering alerts when data is not statistically compatible (accounting for the uncertainties). Furthermore, the uncertainties of the templates serve a dual purpose: they capture acceptable variations in detector response (such as degenerate good solutions where noisy pixels induce visible but acceptable deviations), and when combined with test statistics, they provide the foundation for direct interpretability --- enabling domain experts to verify system outputs and perform root-cause analysis.

In the context of particle physics experiments, a run corresponds to a data-collection interval during which detectors accumulate events. Each run is monitored through a collection of histograms, where the bin counts accumulate information over the whole duration of the run. Therefore, different run durations result in varying total counts and so varying bin-to-bin statistical uncertainties. While we model these uncertainties using Poisson statistics in this work, the DINAMO framework is generalizable to histograms with any type of per-bin uncertainties. 

As discussed in~\autoref{sec:introduction}, the framework comprises two complementary approaches:
\begin{itemize}
\item \textbf{DINAMO-S}: A non-ML, statistics-based approach that extends EWMA with additional statistical re-weighting factors to account for the Poisson uncertainty of good runs. It offers simplicity, scalability, and ease of deployment, potentially enabling rapid adoption in experimental workflows.

\item \textbf{DINAMO-ML}: An ML-based approach that uses a deep neural network that is trained via online learning to predict future references and their uncertainties while taking as input past good histograms. Computationally more intensive than DINAMO-S, it is designed to offer higher accuracy and faster adaptation to changing conditions.
\end{itemize}

We consider a sequence of \emph{runs} indexed by \(i = 1, 2, \ldots, n\). These events can be integrated and visualized via histograms that capture key detector responses (e.g., energy deposits, track multiplicities, etc.) over the course of that run. Runs are ordered sequentially, following the data-collection order. Our proposed algorithms are designed to be applicable in both online and offline regimes of DQM. They assume an operational setup where a trained model quickly produces an anomaly score for each new run. If the score is under a certain threshold, the data can be automatically accepted; otherwise, the system can send the run to human experts, displaying its prediction along with summary plots or statistics. The human shifter can then decide whether the run is genuinely anomalous, or if experimental conditions have changed in a way the algorithm has not yet learned.

While we primarily illustrate the algorithm's behavior using a single histogram per run, most real particle physics applications monitor the response of multiple subsystems, with several histograms measuring different quantities for each subsystem. Extending our approach is straightforward: one can (i) compute a separate anomaly score for each histogram and combine them (e.g., a bin-by-bin \(\chi^2\) aggregated across all relevant histograms) to produce a single run-level score, or (ii) keep multiple scores, per individual histogram or per subsystem (aggregating over the relevant histograms). This flexibility permits granular diagnosis of potential issues, allowing shifters to pinpoint which subsystem may be responsible if the overall run is flagged as anomalous.

Formally, we denote the one-dimensional histogram for run \(i\) by \(\bm{x}_i \in \mathbb{R}^{N_b}\), where \(N_b\) is the number of bins. We associate each run with a label \(y_i \in \{0,1\}\), indicating whether the histogram is good (0) or bad (1). This label may come from a human shifter (in real deployments) or from ground truth (in synthetic experiments). A good label typically implies that the detector output conforms to nominal expectations, whereas a bad label signifies that significant anomalies or faulty conditions are present. Good histograms can vary from run to run due to per-bin statistical and intrinsic uncertainties. Here, the concept of intrinsic uncertainty is meant to cover any type of fluctuation in detector response that is not impacting the subsequent analysis and is hence harmless.

A central concept in our framework is the \emph{reference histogram} \(\bm{\mu}\) and its per-bin uncertainties \(\bm{\sigma_{\mu}}\), which represent the current best estimate of the nominal (i.e., good) distribution for a given type of histogram. When a newly observed histogram \(\bm{x}_i\) is confirmed to be good, the reference is updated to incorporate this information; if a histogram is bad, the reference remains unchanged. Both DINAMO-S and DINAMO-ML produce a \(\chi^2\)-based anomaly score that helps decide if a histogram is consistent with the reference.

\subsection{DINAMO-S: Statistical Method}
\label{subsec:DINAMO_s}

\paragraph{Method description.}
DINAMO-S aims to maintain a reference histogram \(\bm{\mu} \in \mathbb{R}^{N_b}\) (with an associated uncertainty \(\bm{\sigma}_{\mu}\)) that tracks the nominal distribution of the detector subsystem over time. When a new histogram \(\bm{x}_i\) arrives, a test statistic (reduced \(\chi^2\)) is used to determine if \(\bm{x}_i\) is consistent with \(\bm{\mu}\). If \(\bm{x}_i\) is labeled good, the reference is updated via an updated version of EWMA. The extension of EWMA is made to adequately address the varying event statistics and so varying Poisson uncertainty. For that, we introduce an additional re-weighting factor that accounts for the statistical variability of each new run when updating both the reference means and their uncertainties.

\paragraph{Problem setup and notation.}

\begin{itemize}
\item \(\{\bm{x}_1, \bm{x}_2, \ldots, \bm{x}_n\}\) denotes the sequence of \(n\) observed histograms, each with \(N_b\) bins.  
\item Each \(\bm{x}_i\) can be labeled good (\(y_i = 0\)) or bad (\(y_i = 1\)). In real deployments, a human operator (or automated threshold) decides the label; in synthetic tests, the label is taken from ground truth.  
\item The reference histogram \(\bm{\mu}\) is our evolving ``template'' for good data. Its bin-wise uncertainty is \(\bm{\sigma}_{\mu}\).  
\item The unit-normalized version of \(\bm{x}_i\) is denoted as \(\bm{\tilde{x}}_i\), i.e., \(\bm{\tilde{x}}_i\) = \(\bm{x}_i\)/\(I_{x_i}\), where \(I_{x_i}\) is the total integral of \(\bm{x}_i\). Bin-wise Poisson uncertainties are computed accordingly.
\end{itemize}

\paragraph{Reference initialization.}

We initialize \(\bm{\mu}\) as a uniform distribution (e.g., each bin set to a constant value, then normalized to unity). The initial uncertainty \(\bm{\sigma}_{\mu}\) is set to the corresponding Poisson uncertainty per bin. Formally:
\begin{enumerate}
\item Set \(\mu_{0,j} = \rm{const}\) for all bins \(j\); normalize so \(\sum_j \mu_{0,j} = 1\).
\item Compute bin-wise \(\sigma_{\mu_0}\) as \(\sqrt{\mu_{0,j}/I_{\mu_0} - \mu^2_{0,j}/I_{\mu_0}}\), where \(I_{\mu_0}\) is the sum of the original (pre-normalized) values: \(I_{\mu_0} = N_b \cdot \rm{const}\).
\end{enumerate}

\paragraph{EWMA framework.}

To adapt \(\bm{\mu}\) whenever a new histogram is labeled good, we use an EWMA-based scheme with a smoothing factor \(\alpha \in [0,1)\). 
 Internally, we keep track of the following supporting variables or ``accumulators'' (to include the additional re-weighting factor for the statistical uncertainty of a new run):
\[
\bm{W}_i,\quad
\bm{S}_{\mu, i},\quad
\bm{S}_{\sigma_{\mu}, i},
\]
which aggregate weighted statistics. At initialization:
\[
\bm{W}_0 \;=\; (1 - \alpha)\,\bm{\omega}_0,
\quad
\bm{S}_{\mu_0} \;=\; (1-\alpha)\,\bm{\omega}_0\,\bm{\mu}_0,
\quad
\bm{S}_{\sigma_{\mu_0}} \;=\; (1-\alpha)\,\bm{\omega}_0\,\bm{\sigma}_{\mu_0,p}^2,
\]
where \(\bm{\omega}_0\) is a bin-wise weight inversely proportional to \(\bm{\sigma}_{\mu_0,p}^2\).

\paragraph{Test statistic: \(\chi^2\) and pull.}

When a new histogram \(\bm{x}_i\) arrives, we first compute a reduced \(\chi^2\):
\[
\chi^2_\nu
\;=\;
\frac{1}{N_b}
\sum_{j=1}^{N_b}
\frac{\bigl(\tilde{x}_{i,j} - \mu_{i, j}\bigr)^2}
     {\sigma_{\tilde{x}_{i,j},p}^2 + \sigma_{\mu_{i, j}}^2}.
\]
The pull $\bm{\delta}_i$ provides bin-wise deviations:
\[
\delta_{i,j}
\;=\;
\frac{\tilde{x}_{i,j} - \mu_{i,j}}
     {\sqrt{\sigma_{\tilde{x}_{i,j},p}^2 + \sigma_{\mu_{i,j}}^2}}.
\]
Although both \(\chi^2_\nu\) and \(\bm{\delta_{i}}\) aid human shifters in deciding whether the run is indeed good or bad, the algorithm's decision is based solely on the reduced \(\chi^2_\nu\) value compared to a predetermined threshold. The pull vector \(\bm{\delta_{i}}\) serves as a tool that aids human operators in understanding which specific bins contribute most to the anomaly, enabling efficient root-cause analysis and validation of the algorithm's decision.

\paragraph{Reference update.}

If the run ${\bm{{x}}_i}$ is confirmed good (\(y_i = 0\)), we update \(\bm{\mu}\) and \(\bm{\sigma}_\mu\) bin by bin. First, we normalize ${\bm{{x}}_i}$ to unity and obtain $\bm{\tilde{x}}_i$: $\bm{\tilde{x}}_i = \bm{x}_i \ / \ I_{x_i}$, where $I_{x_i}$ is a sum across all bins $N_b$: $I_{x_i} = \sum_{j}^{N_b}{x_{i, j}}$. Then, let the statistical uncertainty factor $\bm{\omega}_i$ of the new run, used for the reference update, be:
\[
\bm{\omega}_i
\;=\;
\frac{1}{\,\bm{\sigma}^2_{\bm{\tilde{x}}_i, p} \;+\;\varepsilon\,}, 
\]
where \(\bm{\sigma}_{\bm{\tilde{x}}_i, p} = \sqrt{\frac{\bm{\tilde{x}}_i}{I_{x_i}} - \frac{\bm{\tilde{x}}^2_i}{I_{x_i}}}\) is the bin-wise estimation of the Poisson uncertainty for $\bm{\tilde{x}}_i$, and $\varepsilon$ is a small value.
Then we accumulate:
\begin{eqnarray*}
    \bm{W}_{i+1} &\;=\; \alpha\,\bm{W}_{i}
    \;+\;
    (1-\alpha)\,\bm{\omega}_i,
    \\
    \bm{S}_{\mu, i+1}
    &\;=\;
    \alpha\,\bm{S}_{\mu, i}
    \;+\;
    (1-\alpha)\,\bm{\omega}_i\,\bm{\tilde{x}}_i,
    \\
    \bm{S}_{\sigma_\mu, i+1}
    &\;=\;
    \alpha \,\bm{S}_{\sigma_\mu, i}
    \;+\;
    (1-\alpha)\,\bm{\omega}_i \,\bigl(\bm{\tilde{x}}_i - \bm{\mu}_i\bigr)^2,
    \\
    \bm{\mu}_{i+1}
    &\;=\;
    \frac{\bm{S}_{\mu, i+1}}{\bm{W}_{i+1}};
    \ \bm{\sigma}_{\bm{\mu}, i+1}
    \;=\;
    \sqrt{\frac{\bm{S}_{\sigma_\mu, i+1}}{\bm{W}_{i+1}}}.
\end{eqnarray*}

The computed uncertainty effectively absorbs a contribution from the limited statistical knowledge of the reference, in addition to the intrinsic uncertainty. However, that statistical contribution is very small in practice. The full pseudocode of the DINAMO-S method is provided in~\autoref{alg:dinamo_s} in~\ref{app:details_dinamo_s}.

\subsection{DINAMO-ML: Machine-Learning Method}
\label{subsec:DINAMO_ml}

\paragraph{Method overview.}
DINAMO-ML replaces the hand-crafted EWMA update mechanism with a learnable transformation from the last \(M\) good histograms to a predicted reference \(\boldsymbol{\hat{\mu}}\) and its uncertainty \(\boldsymbol{\hat{\sigma}}\). Both \(\boldsymbol{\hat{\mu}}\) and \(\boldsymbol{\hat{\sigma}}\) are inferred via a Transformer encoder-based network that processes these \(M\) previous good histograms along with their relative timestamps (where ``relative'' means ``with respect to the current $i$-th run''). The algorithm continues to produce a \(\chi^2\) and pull $\delta$, mirroring DINAMO-S in interpretability and overall workflow (predict-reference, compute-anomaly-score, decide-good-or-bad, update-if-good), but it learns the temporal weighting instead of relying on a single smoothing factor \(\alpha\). In practice, we set the anomaly threshold using historical data (e.g., a validation sample of 20\% of labeled runs at the beginning of data collection), optimizing for an acceptable trade-off between false positives and false negatives.

\begin{figure*}[!htb]
\begin{center}
\centerline{\includegraphics[width=\textwidth]{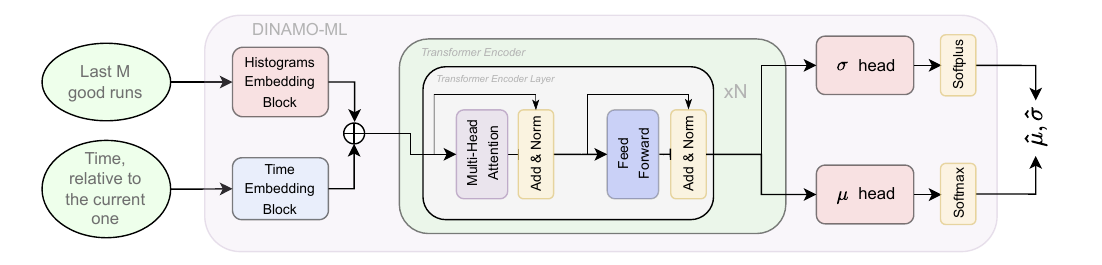}}
\caption{A schematic representation of the Transformer encoder-based DINAMO-ML model. The input embeddings and output heads are implemented using a simple fully connected block: \texttt{{Linear, LayerNorm, ReLU, Linear}}. To ensure the positivity of uncertainty values, the \texttt{Softplus} activation function is applied, while \texttt{Softmax} is used for the reference $\hat{\mu}$ to enforce normalization to unity.}
\label{fig:dinamo_ml_scheme}
\end{center}
\end{figure*}

\paragraph{Architecture and reference prediction.}
Whereas DINAMO-S uses a hand-crafted EWMA update for the reference's $\bm{\mu}$ and $\bm{\sigma_{\mu}}$, the DINAMO-ML approach learns an adaptive mapping from past good histograms to the current reference. Specifically, a Transformer encoder-based network takes as input the last $M$ good histograms (additionally with their relative time to run $i$, counted in terms of number of runs) and produces per-bin outputs $(\bm{\hat{\mu}}, \bm{\hat{\sigma}})$. A schematic view of the model, implemented using the \texttt{TransformerEncoder} class from PyTorch~\cite{10.5555/3454287.3455008}, is shown in \autoref{fig:dinamo_ml_scheme}. 

At each \(i\)-th run, let \(\bm{G}_i\) be the set of the last \(M\) good histograms (with timestamps less than $i$). These are encoded as a sequence of ``tokens'' for the Transformer, which outputs \(\hat{\mu} \in \mathbb{R}^{N_b}\) and \(\hat{\sigma} \in \mathbb{R}^{N_b}\). These serve as the predicted reference histogram and its bin-wise uncertainty, respectively. If there are fewer than \(M\) good runs available (e.g., at the start of data-taking), we pad the context with a constant to maintain consistent input shapes.

The output of the transformer is then used to construct a \emph{conditional Probability Density Function (PDF)} that models what a good histogram at time $i$ is expected to look like. Concretely, each bin's content $\tilde{x}_{i,j}$ is modeled as a Gaussian random variable with mean $\hat{\mu}_j$ and variance $\hat{\sigma}_j^2$. The conditional PDF hence looks like:
\[
p(\boldsymbol{\tilde{x}}_i \mid \{\boldsymbol{\tilde{x}}_{g}\}_{g\in \mathcal{G}_i})
\;=\;
\prod_{j=1}^{N_b}
\mathcal{N}\Bigl(\tilde{x}_{i,j}\;\Big|\;\hat{\mu}_j,\;\hat{\sigma}_j^2\Bigr),
\]
where $\mathcal{G}_i$ denotes the indices of the last $M$ good histograms preceding the $i$-th run\footnote{In practice, if fewer than $M$ good histograms are available (e.g.\ in early runs), the context is constant-padded.} and $\mathcal{N}$ denotes a Gaussian distribution.

\paragraph{Training procedure.}
From a probabilistic standpoint, DINAMO-ML performs a mini-batch likelihood fit over good histograms. Minimizing the Gaussian Negative Log-Likelihood (NLL) $\mathcal{L}$ of good runs under this PDF, via mini-batch training, encourages $(\boldsymbol{\hat{\mu}}, \boldsymbol{\hat{\sigma}})$ to capture the temporal and bin-to-bin dependencies:
\[
\mathcal{L}
=
\frac{1}{2N_bK}
\sum_{k=1}^{K}
\sum_{j=1}^{N_b}
\biggl[
  \left(\frac{\tilde{x}_{k,j} - \hat{\mu}_{k, j}}{\hat{\sigma}_{k, j}}\right)^2
  + \log{(\hat{\sigma}_{k, j}^2)}
\biggr],
\]
where $K$ is the mini-batch size and $\hat{\mu}_k$ and $\hat{\sigma}_k$ are computed by the network from the $M$ good histograms preceding run $k$.
This likelihood-minimization analogy holds exactly only for the current mini-batch; the trained model is not constrained to describe past histograms outside this window. It is important to note that for each mini-batch, the model is trained iteratively on the same batch until convergence (monitored via early stopping), rather than making a single gradient update.

To implement mini-batch updates in practice:
\begin{enumerate}
  \item We additively create a dataset \(\mathcal{D}\) by collecting good runs and their timestamps as they become available.
  \item After the new good run added, we gather a mini-batch of \(K\) chronologically recent good histograms from \(\mathcal{D}\). If fewer than \(K\) exist, we take as many as available.
  \item For each run in this batch, we reconstruct the context of the last \(M\) good histograms preceding it (constant-padding if needed), run the network to get \((\hat{\mu}_k, \hat{\sigma}_k)\), and compute the NLL loss above.
  \item We backpropagate to adjust the Transformer encoder's parameters, effectively learning how to predict the bin-by-bin means and uncertainties for future runs.
  \item At each reference update step (so, training step), we use the early stopping technique with a patience of 5, terminating if the loss fails to improve further. 
\end{enumerate}

\paragraph{Inference and anomaly detection.}
The process for a new run follows these steps:
\begin{enumerate}
\item \textbf{Context Construction:} For a new run \(\boldsymbol{x}_i\), we identify up to \(M\) most recent good histograms (with time $<$ \(i\)) to form a conditional context (again constant-padding if fewer than \(M\) exist). 
\item \textbf{Predict Reference:} Process the set of $M$ histograms and their relative time, obtaining the transformer encoder model outputs \(\hat{\boldsymbol{\mu}}_i, \hat{\boldsymbol{\sigma}}_i\).
\item \textbf{Anomaly Score:} Then, we compute
\[
\chi^2_\nu
\;=\;
\frac{1}{N_b}
\sum_{j=1}^{N_b}
\frac{\bigl(\tilde{x}_{i,j} - \hat{\mu}_{i, j}\bigr)^2}
     {\sigma_{\tilde{x}_{i,j},p}^2 + \hat{\sigma}_{i, j}^2}
\]
and a pull vector $\boldsymbol{\delta}_i$:
\[
\delta_{i,j}
\;=\;
\frac{\tilde{x}_{i,j} - \hat{\mu}_{i,j}}
     {\sqrt{\sigma_{\tilde{x}_{i,j},p}^2 + \hat{\sigma}_{i, j}^2}}.
\]
Here, $\boldsymbol{\tilde{x}}_i$ is the run's histogram normalized to unity, so that 
\[
\boldsymbol{\tilde{x}}_i = \boldsymbol{x}_i / I_{x_i},
\]
where $I_{x_i} = \sum_{j=1}^{N_b} x_{i,j}$, and
\[
\boldsymbol{\sigma}_{\boldsymbol{\tilde{x}}_i, p} = \sqrt{\boldsymbol{\tilde{x}}_i / I_{x_i} - \boldsymbol{\tilde{x}}^2_i / I_{x_i}}
\]
is the bin-wise estimation of the Poisson uncertainty for $\boldsymbol{\tilde{x}}_i$. To decide whether \(\boldsymbol{x}_i\) is good or bad, the anomaly score is compared to a predetermined threshold.

\item \textbf{Update:} If the $i$-th run is labeled good, we add \((\boldsymbol{\tilde{x}}_i, i)\) to the dataset $\mathcal{D}$; if bad, we exclude it.
\end{enumerate}

Compared to the statistical method, DINAMO-ML can learn more flexible temporal ``kernels'' to weight or combine previous good histograms, potentially adapting more quickly to abrupt shape changes. The full pseudocode of the DINAMO-ML method is provided in~\autoref{alg:dinamo_ml} in~\ref{app:details_dinamo_ml}.

\section{Goals and Metrics}
\label{sec:goals_metrics}

In this section, we describe the key metrics we use to evaluate the performance of both DINAMO-S and DINAMO-ML. While standard anomaly detection often centers on classification metrics alone, our approach must also handle evolving conditions and provide interpretable templates and their uncertainties. Consequently, we focus on the following three metrics:

\paragraph{Balanced Accuracy.}
Because the fraction of good and bad histograms can deviate from a 50--50 split and vary per dataset, we rely on balanced accuracy to assess classification performance. Balanced accuracy is computed as the average of recall over each class (good/bad), mitigating potential bias in the presence of class imbalance. In our binary classification case, this equals the average of sensitivity (true positive rate) and specificity (true negative rate). To further evaluate classification quality on a per-class basis, we also report sensitivity and specificity themselves.

\paragraph{Adaptation time.}
We quantify how quickly an algorithm recovers when operating conditions change. Specifically, we count how many subsequent good runs are misclassified (i.e., have $\chi^2$ above the threshold) before the algorithm adapts and correctly classifies good runs again. We then average this count across all rapid changes in the dataset, defining a measure of adaptation time. Therefore, this metric shows an average amount of good runs to be misclassified before an algorithm adapts to the new conditions. If a good run falls below the threshold immediately after the change, the adaptation metric yields zero, which limits it to the range of [0, +$\infty$].

\paragraph{Uncertainty Coverage.}
To assess how well the algorithms' bin-wise uncertainty estimates reflect true variations in the data, we computed the \emph{Jaccard distance}\footnote{The Jaccard distance is defined as $D_J = 1 - \rm{IoU}$ where IoU is the Intersection over Union. Note that the Jaccard distance is a metric (by mathematical definition)~\cite{KOSUB201936}.} between the predicted uncertainty distributions and the empirically observed distributions of good runs. Since data evolve over time, a direct comparison is challenging. To address this, we first standardize each good run histogram by converting it into z-score space.\footnote{Specifically, for each histogram, we transform the bin centers using $\mu_{\mathcal{N}}$ and $\sigma_{\mathcal{N}}$ of the corresponding Gaussian distribution: $z = (u - \mu_{\mathcal{N}}) \ / \ \sigma_{\mathcal{N}}$. This standardization ensures that the histograms are comparable across different runs.} Once standardized, we estimate the probability density function by normalizing the counts relative to the transformed bin widths. The same transformation is then applied to the model's predicted references and their uncertainty estimates. Finally, to facilitate a consistent comparison between the model's output and the true variability of the data, we interpolate all transformed distributions onto a common set of bin centers. The Jaccard distance is then computed by comparing the geometric areas covered by the 1-$\sigma$ uncertainty bands: $D_J = 1 - \frac{A_{\rm{intersection}}}{A_{\rm{union}}}$, where $A_{\rm{intersection}}$ is the overlapping area between the predicted and empirical uncertainty bands, and $A_{\rm{union}}$ is the total area covered by either band. A Jaccard distance of zero indicates perfect alignment between the model's uncertainty estimates and the empirical distribution, while a Jaccard distance of one signifies complete misalignment.

\section{Synthetic data description}
\label{sec:data}

To systematically evaluate and compare the DINAMO algorithms, a controlled testing environment with well-defined properties is needed. This section describes the synthetic data generation process that was designed to mimic real-world conditions while providing ground truth labels for validation. Each run in our synthetic dataset produces a one-dimensional histogram drawn from a Gaussian distribution with a varying number of events, whose parameters can drift over time or change abruptly. In addition, a systematic uncertainty modeling is performed to bring it closer to real-world examples.
Summary of the main features of the data generation can be found below:
\begin{itemize}
    \item \textbf{Slow operational drifts.} The mean of the Gaussian distribution may evolve gradually (e.g., sinusoidally), reflecting slow changes in detector or hardware conditions.
    \item \textbf{Abrupt shifts.} With a certain probability, sudden jumps occur in the mean or width of the Gaussian distribution, modeling issues like hardware resets or configuration updates.
    \item \textbf{Varying event statistics.} Initial total number of events of each run is drawn uniformly from a range of values, mimicking fluctuations in data collection rates and introducing different Poisson uncertainties.
    
    \item \textbf{Systematic uncertainty modeling.} On top of the Poisson uncertainty, a systematic binomial-like uncertainty is introduced. For each run, the bin contents in the right half of the histogram are modified by adding a random fluctuation. Specifically, for each bin $b$ in the right half, the bin content $n_b$ is modified to $n_b \pm \Delta n_b$, where $\Delta n_b$ is sampled  from a binomial distribution $B(n_b, p)$. Here, $p$ is a pre-defined parameter that controls the magnitude of systematic fluctuations. The sign (+ or -) is chosen randomly for each run to model either accidental increases or decreases of events, mimicking correlated inefficiencies in readout electronics of the detector.
    \item \textbf{Bad runs: extra distortions.} A subset of all the runs may exhibit additional parameter shifts or partially missing bins with a certain probability. These runs are labeled as bad.
\end{itemize}

While the development of the synthetic data generator was motivated by our experience with the LHCb experiment, the generator is designed to represent general characteristics common across particle physics experiments rather than specific LHCb operational parameters.

A Python implementation of this generator provides a straightforward interface for specifying parameter values that control all above-mentioned features, seeding the randomness and saving each dataset for subsequent experiments. Therefore, by adjusting the parameters, we can create relatively simple datasets (few rapid changes, smaller amount of bad runs, etc.) or more challenging ones (frequent abrupt shifts, bad runs domination, etc.).

In this study, we generated 1000 synthetic datasets with a certain parameter setting (e.g., the fraction of anomalous runs, magnitude of abrupt shifts) but varying random seeds. This helps quantify the performance of both models on many datasets and statistically compare them. Each dataset consists of 5000 runs, with 500 of them being bad runs.~\ref{app:synthetic_data_desc} provides more technical details, the full pseudocode, and the exact parameter settings used in the study.

\begin{figure}[h]
\vskip 0.2in
\begin{center}
\centerline{\includegraphics[width=\textwidth]{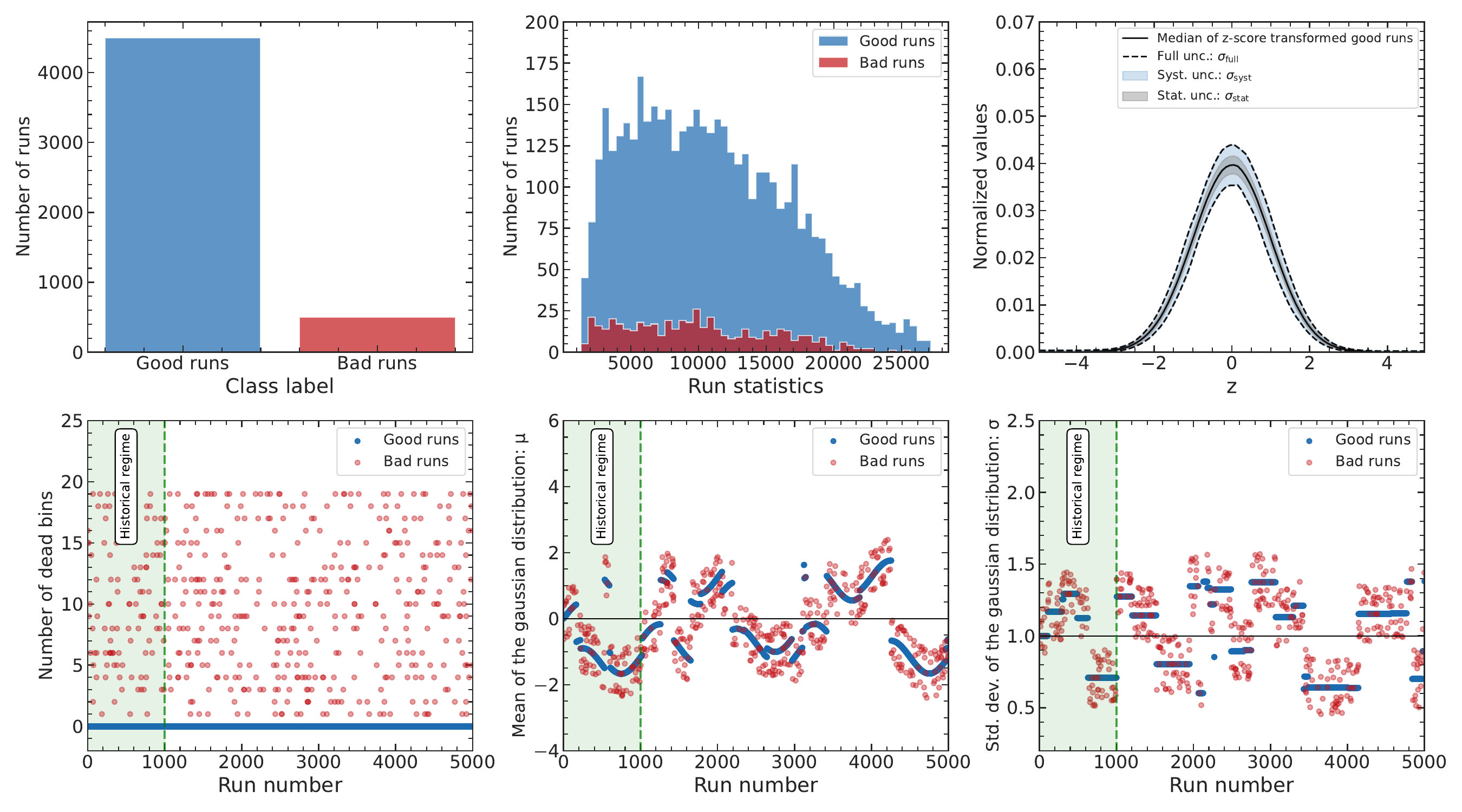}}
\caption{General overview of a single dataset produced using the data generator. Top row shows class label distribution, final runs statistics distribution and the uncertainty visualization in the z-score transformed space. The bottom row presents the time evolution of the parameters of the Gaussian distributions and the number of dead bins for the bad runs as a function of run number.}
\label{fig:dataset_description}
\end{center}
\vskip -0.2in
\end{figure}

\autoref{fig:dataset_description} shows the main distributions characterizing
one of the generated datasets with a random seed of 70. The top row of the~\autoref{fig:dataset_description} illustrates the class distribution, final runs statistics and the uncertainty visualization (in the z-score transformed space). The bottom row depicts the time evolution of both the parameters of the Gaussian distributions and the number of dead bins for the bad runs as a function of run number. The set of 1000 datasets is used in~\autoref{sec:results} to assess the performance of the DINAMO models.

\section{Results}
\label{sec:results}

In this study, each dataset is split into a \emph{historical} regime (the first 1000 runs, 20\%) used for hyperparameter tuning and threshold optimization, and a \emph{continual} regime (the remaining 4000 runs) for performance evaluation. Using the historical data, we tune the optimal $\alpha$ value in terms of the ROC-AUC score for the DINAMO-S algorithm per each dataset. Instead, DINAMO-ML has a higher number of hyperparameters that define the model: the context buffer $M$, the mini-batch size $K$, the learning rate, and the architecture-related parameters (e.g. number of encoder layers, number of heads, number of expected input features for the encoder layers, dimension of the feedforward network model in the encoder layers, etc.). Using a trial-and-error approach, the following effective hyperparameters were selected:
\begin{itemize}
  \item \(M=20\) for the context buffer (the last 20 good runs).
  \item \(K=10\) for the mini-batch size.
  \item Transformer encoder hyperparameters: 
   the number of encoder layers \(n_{\rm layers}=3\), 
  the embedding dimension \(d_{\rm model}=100\),
  the number of attention heads \(n_{\rm head}=10\), 
  the feed-forward dimension \(d_{\rm ff}=100\),
  the dropout rate \(p_{\rm drop}=0.15\), and the \texttt{GeLU} activation function~\cite{gelu} .
  \item Learning rate of \(5\times10^{-4}\) of the AdamW optimizer~\cite{adamw} with the $L_2$ regularization strength (weight decay) of \(10^{-4}\).
  \item Early stopping with a patience of 5 based on the Gaussian NLL loss $\mathcal{L}$.
\end{itemize}

\begin{figure*}[!htb]
    \centering
    \includegraphics[width=0.49\textwidth]{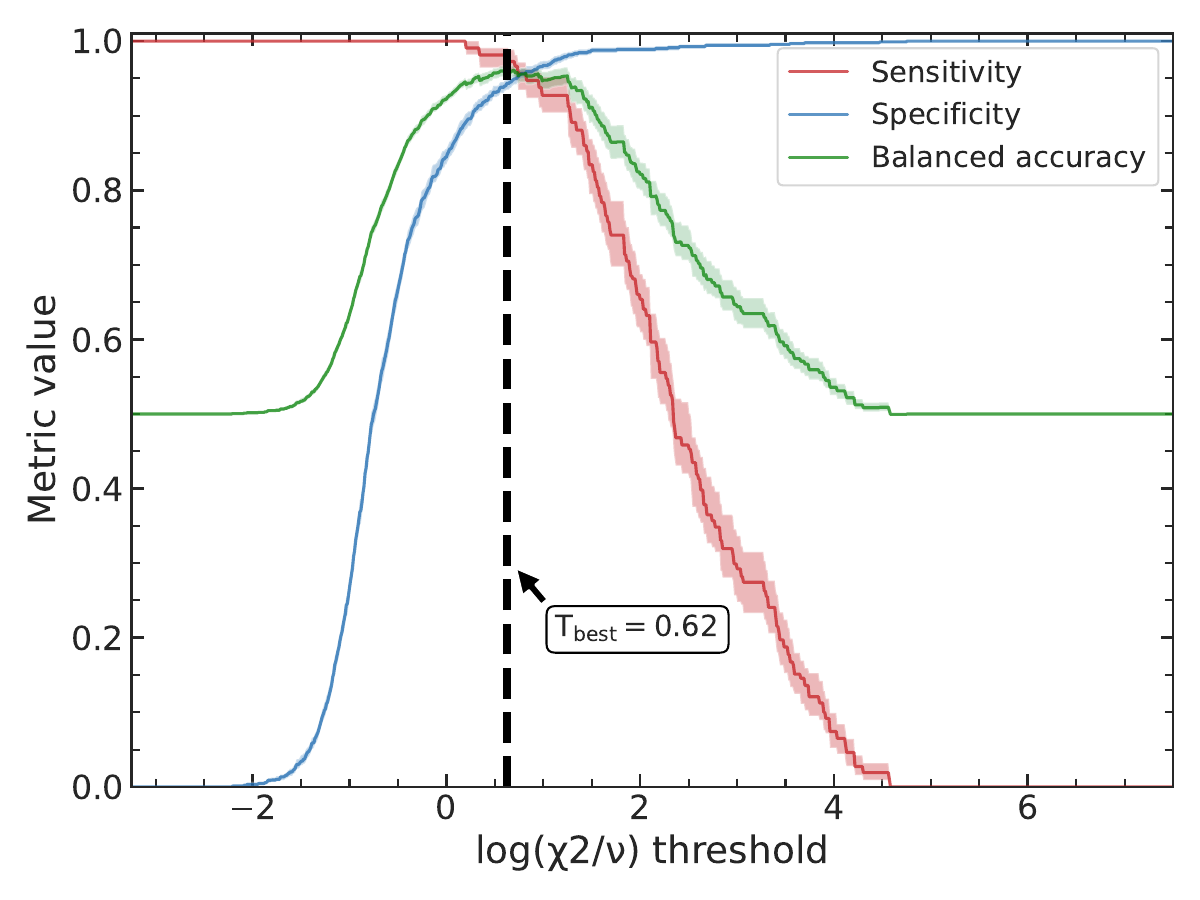}
    \includegraphics[width=0.49\textwidth]{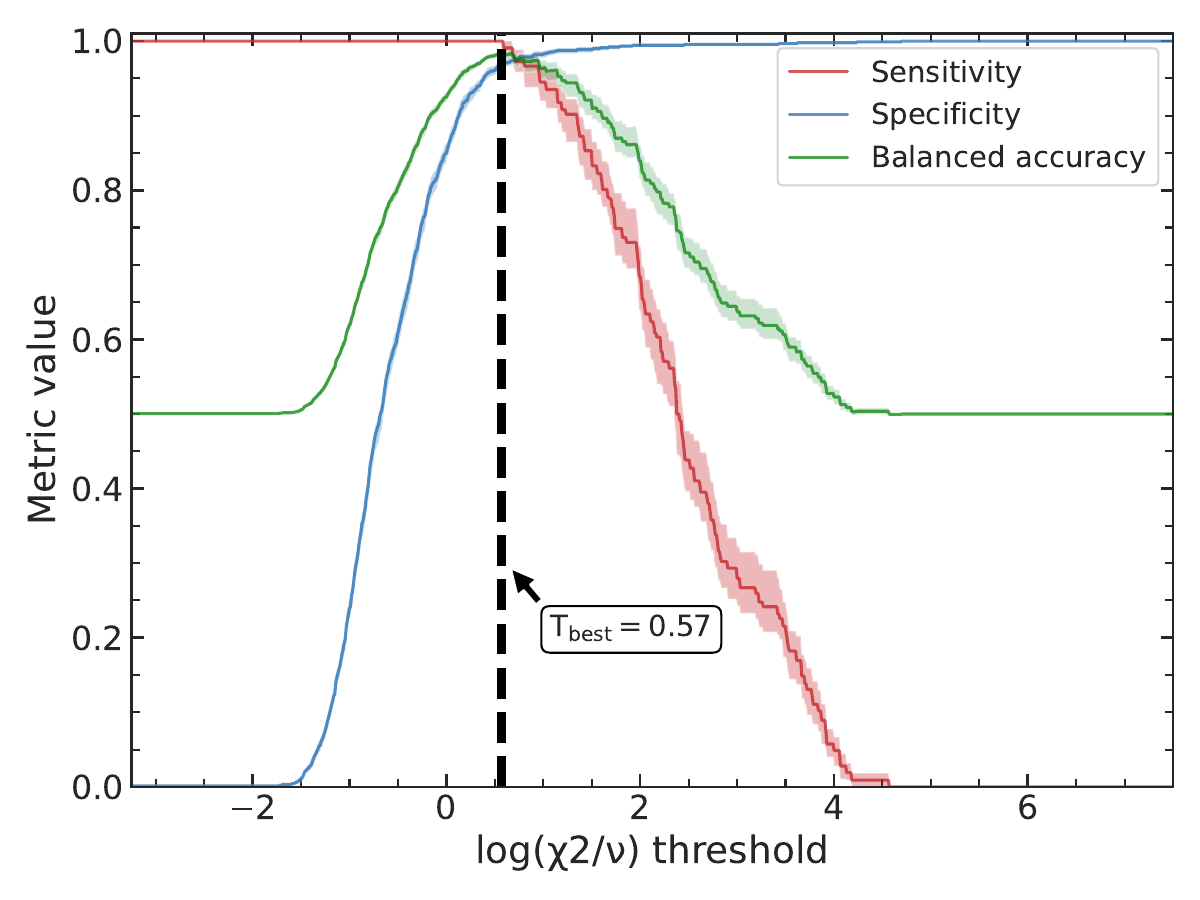}
    \caption{Example of threshold optimization procedure for both DINAMO-S (left) and DINAMO-ML (right) models on the same dataset generated with the seed of 70.}
    \label{fig:best_threshold}
\end{figure*}

Furthermore, we optimize the threshold between classes based on the historical regime data, maximizing balanced accuracy for both algorithms. Finally, the models' performances are then evaluated using only continual regime runs.~\autoref{fig:best_threshold} shows an example of such threshold optimization for both DINAMO-S (left) and DINAMO-ML (right) models on the same dataset generated with the seed of 70.

\paragraph{Results on a Single Synthetic Dataset.}

\begin{figure*}[!htb]
\begin{center}
\includegraphics[width=1\textwidth]{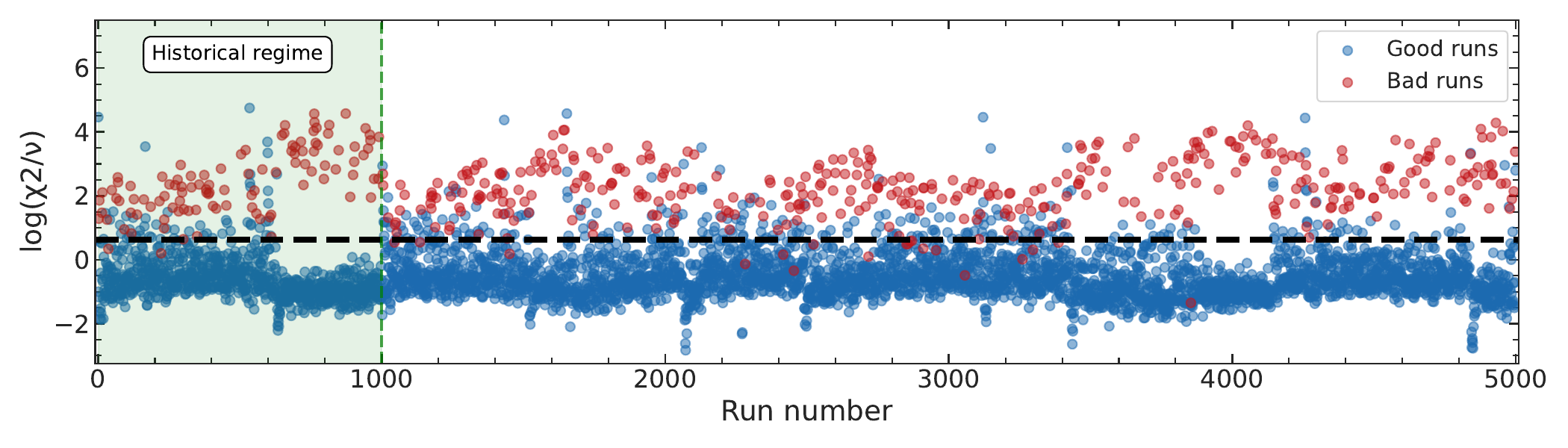}
\caption{Anomaly score of the DINAMO-S algorithms as a function of the run number for a single dataset composed of 5000 runs. Red points represent the ground-truth bad runs and the blue points represent the good runs. The black dashed line is the optimized, using the historical regime, threshold to assign a class label: if a run is above it, the run is predicted to be bad and vice versa.}
\label{fig:dinamo_s_scatter_plot}
\end{center}
\end{figure*}

To illustrate the performance of the DINAMO models on a representative example, we first focus on the same single synthetic dataset with a random seed of 70, mentioned above, in which both slow drifts and rapid changes are injected.~\autoref{fig:dinamo_s_scatter_plot} shows the logarithm of the reduced $\chi^2$ anomaly score for each run for the DINAMO-S model, highlighting how changes in conditions manifest as temporary elevations in the score --- even for runs that are ultimately labeled good. In particular, runs after a condition shift can exceed the chosen anomaly threshold, only to be correctly classified as good once the reference template adapts. Note that the black dashed horizontal line is the optimized, on the historical regime data, threshold value. Analogous results for the ML version of the algorithm can be found in~\ref{app:details_dinamo_ml}.

\begin{figure*}[!htb]
    \centering
    \includegraphics[width=0.49\textwidth]{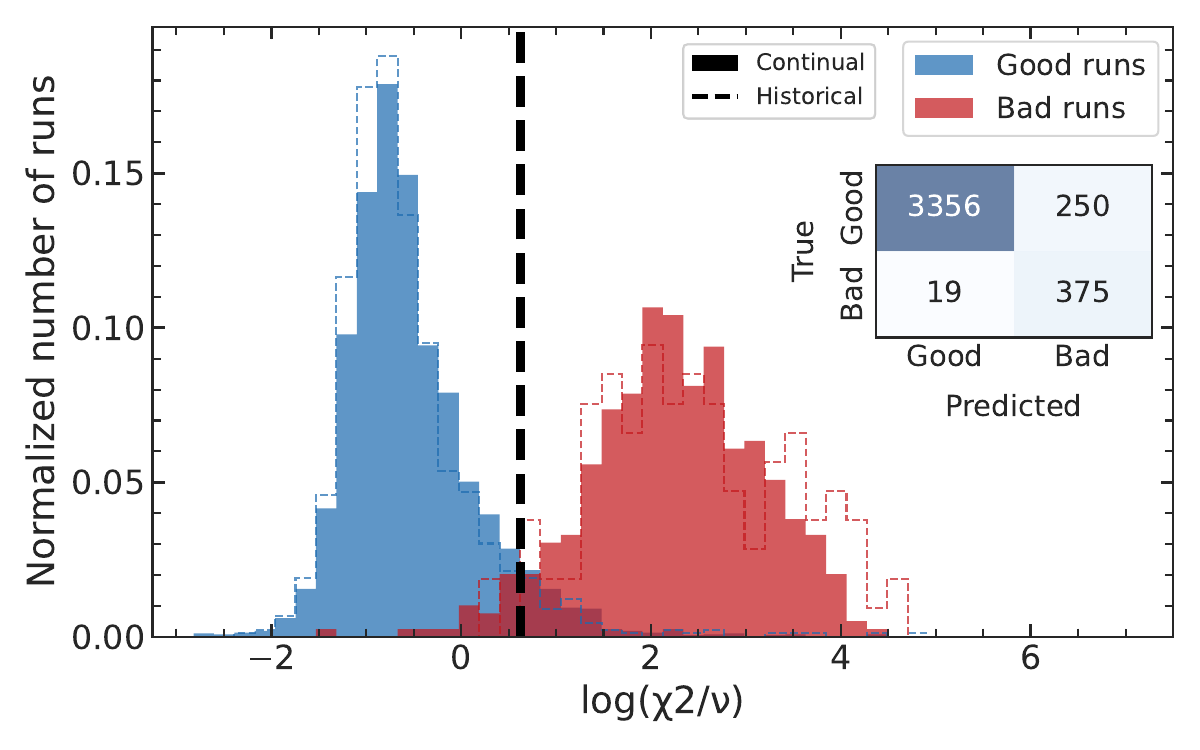}
    \includegraphics[width=0.49\textwidth]{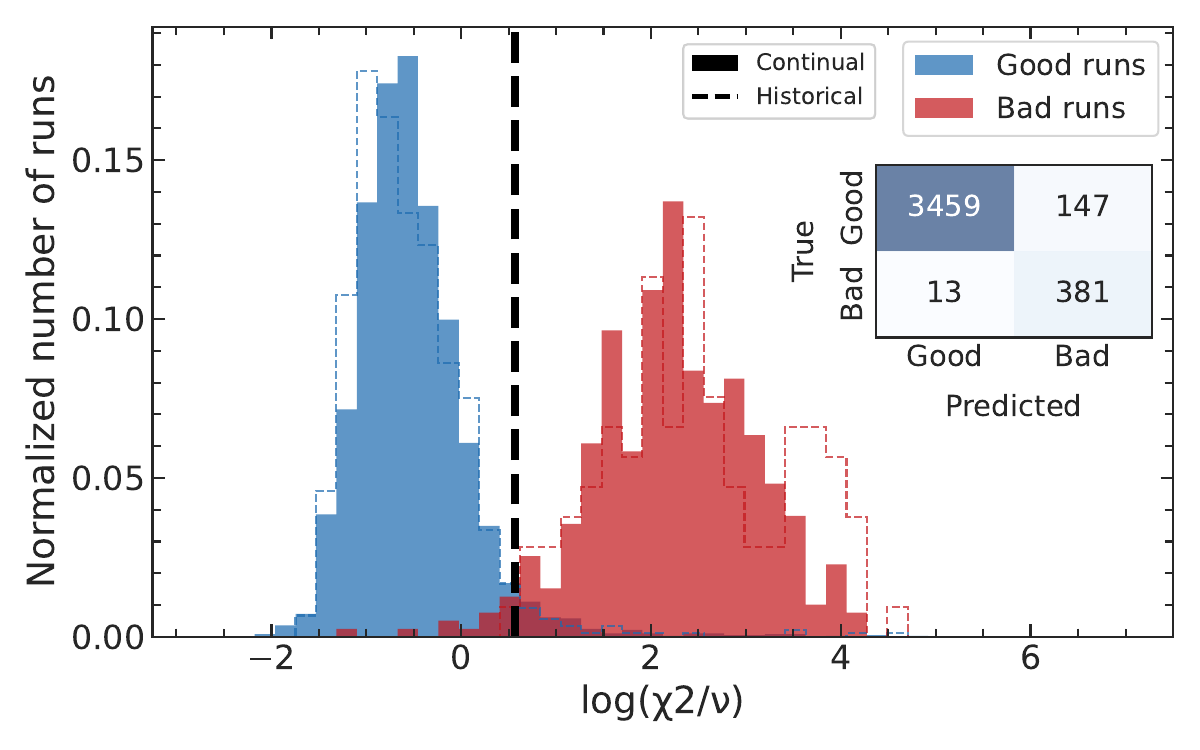}
    \caption{Distributions of the logarithm of the reduced $\chi^2$ anomaly score for each run for DINAMO-S (left) and DINAMO-ML (right) approaches. Filled histograms represent results in the continual regime, while the dashed lines correspond to the historical regime only.}
    \label{fig:results_hist}
\end{figure*}

The models' performance can also be evaluated by examining the distributions of the anomaly scores.~\autoref{fig:results_hist} shows such distributions for both DINAMO-S (left) and DINAMO-ML (right) approaches. The models show low misclassification rates, and the results are consistent across both the historical and continual regimes.

\begin{figure*}[!htb]
	\centering
    \includegraphics[width=\textwidth]{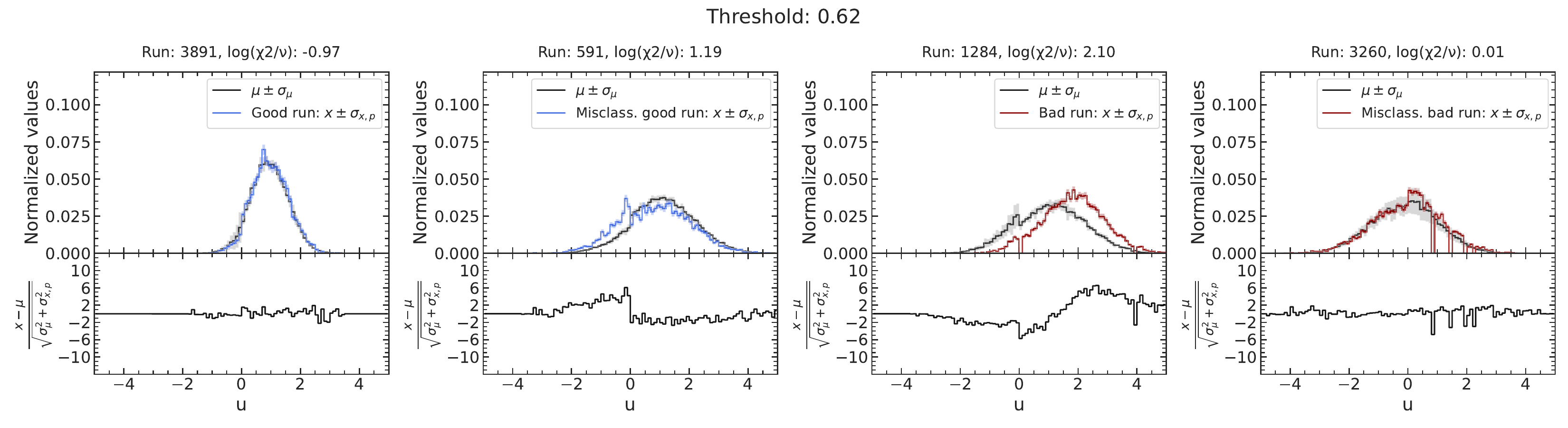}
    \includegraphics[width=\textwidth]{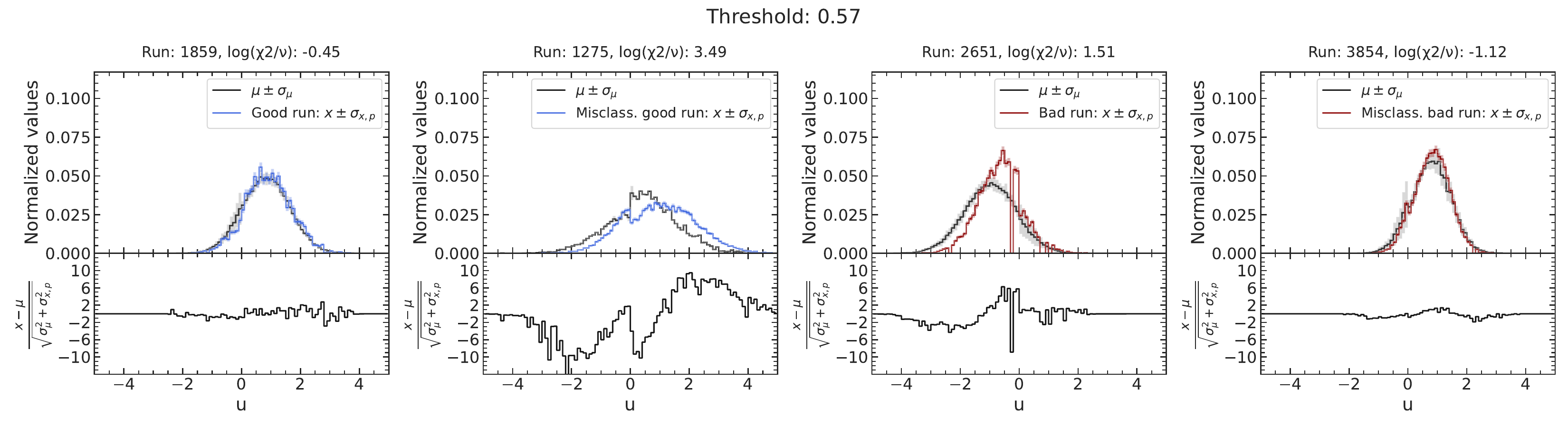}
    \caption{Examples of histograms and references for DINAMO-S (top row) and DINAMO-ML (bottom row). The four following cases are present: 1) a good run correctly classified; 2) a good run incorrectly classified; 3) a bad run, correctly classified; 4) a bad run, incorrectly classified. The corresponding references and their uncertainties are depicted by the black line and the grey shape, respectively. Each plot is also accomplished with a bottom row that represents the pull values.}
	\label{fig:ref_examples_dinamo_s}
\end{figure*}

A closer look at the run-by-run level, can help to demonstrate the models' outputs better. ~\autoref{fig:ref_examples_dinamo_s} provides examples of histograms and references for DINAMO-S (top row) and DINAMO-ML (bottom row). The four following cases are present: 1) a good run correctly classified; 2) a good run incorrectly classified; 3) a bad run, correctly classified; 4) a bad run, incorrectly classified.
Thus, beyond numerical anomaly scores, DINAMO algorithms produce evolving templates and their corresponding uncertainties to help humans quickly assess the quality of the run.

\begin{figure*}[!htb]
\centering
\includegraphics[width=0.49\textwidth]{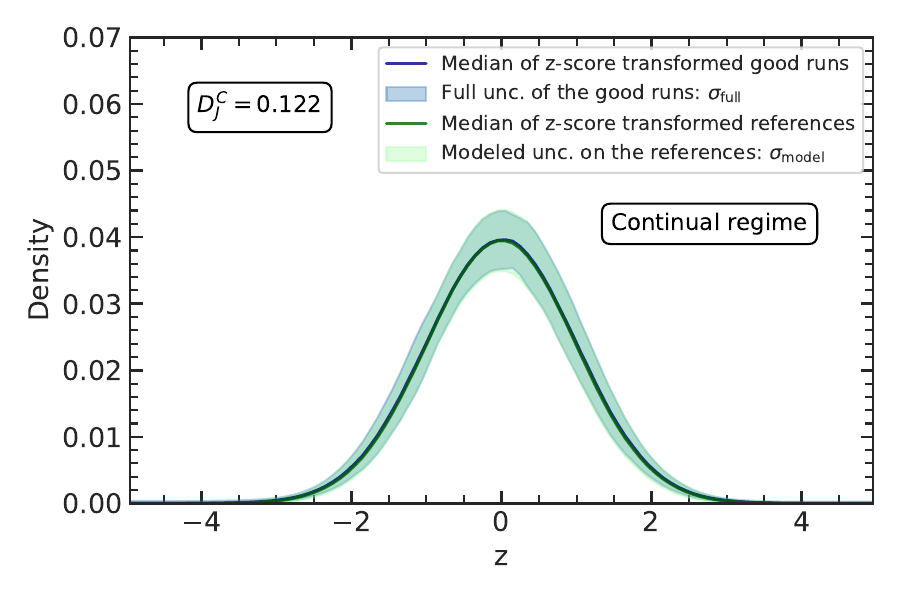}
\includegraphics[width=0.49\textwidth]{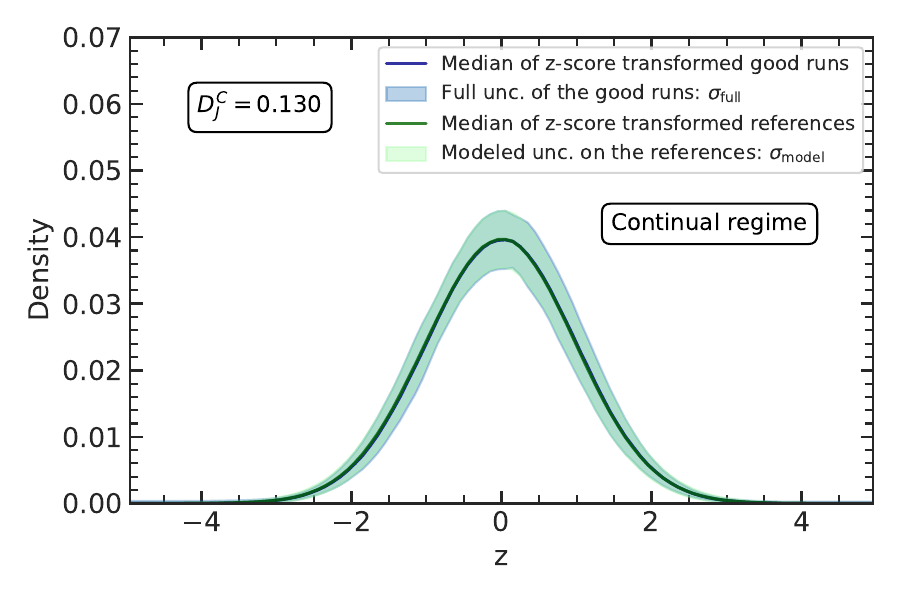}
\caption{Comparison of the references with their uncertainties produced by the DINAMO-S (left) and DINAMO-ML (right) algorithms and corresponding distribution of good run histograms (where both shown in the z-score transformed space). As a measure between the two areas, the Jaccard distance is adopted.}
\label{fig:dinamo_uncertainty}
\end{figure*}

Finally,~\autoref{fig:dinamo_uncertainty} demonstrates the extent to which the predicted templates (with uncertainties) cover the good histogram distributions in this dataset, underscoring the algorithm's ability to track natural fluctuations while still detecting genuine anomalies.

\paragraph{Aggregated Results on 1000 Synthetic Datasets.}

\begin{figure*}[!htb]
    \centering
    \includegraphics[width=0.49\textwidth]{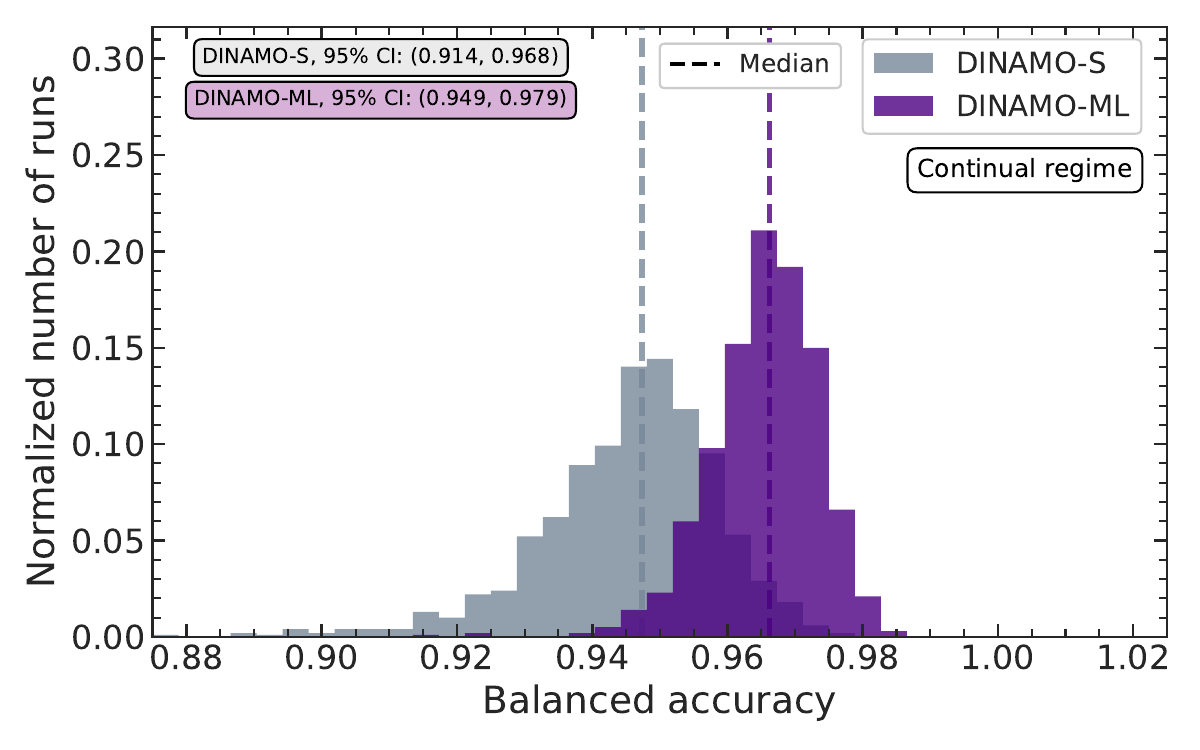}
    \includegraphics[width=0.49\textwidth]{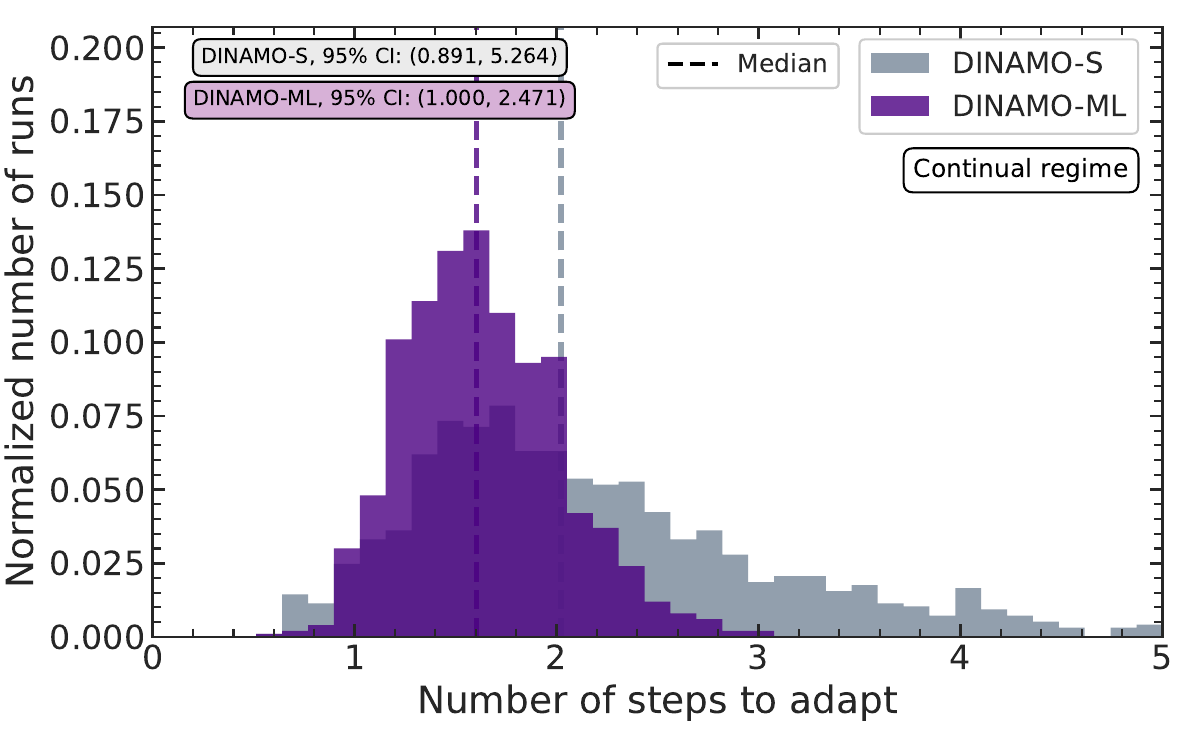}
    \caption{Distributions of balanced accuracies (left) and adaptation time (right; in number of run steps before the convergence) computed over 1000 different synthetic datasets for both DINAMO-S (in grey) and DINAMO-ML (in violet) methods. The dashed lines represent the medians of the distributions. The metrics are computed in the continual regime only.}
    \label{fig:models_comparison_ba_at}
\end{figure*}

To obtain reliable quality metrics and statistically compare the two methods, we generated 1000 datasets with different random seeds while keeping the dataset generator parameters unchanged.~\autoref{fig:models_comparison_ba_at} presents a comparison between DINAMO-S and DINAMO-ML models for the resulting balanced accuracy (left plot) and adaptation time (right plot), computed for runs in the continual regime. While both methods demonstrate high accuracy and adaptability, the ML version statistically outperforms the non-ML one in both metrics. Specifically, with a median value of 0.966 (vs. 0.947 for DINAMO-S) for balanced accuracy and a median value of 1.61 (vs. 2.02 for DINAMO-S) for adaptation time.

\begin{figure*}[!htb]
\centering
\includegraphics[width=0.49\textwidth]{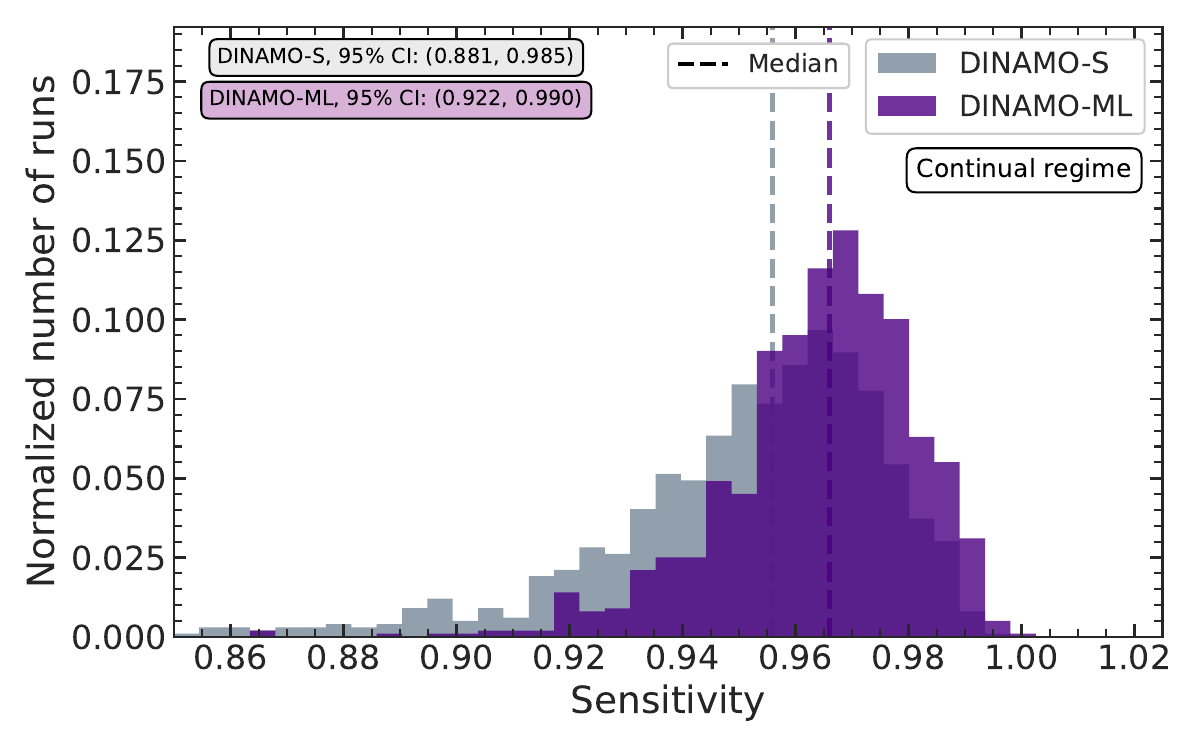}
\hfill
\includegraphics[width=0.49\textwidth]{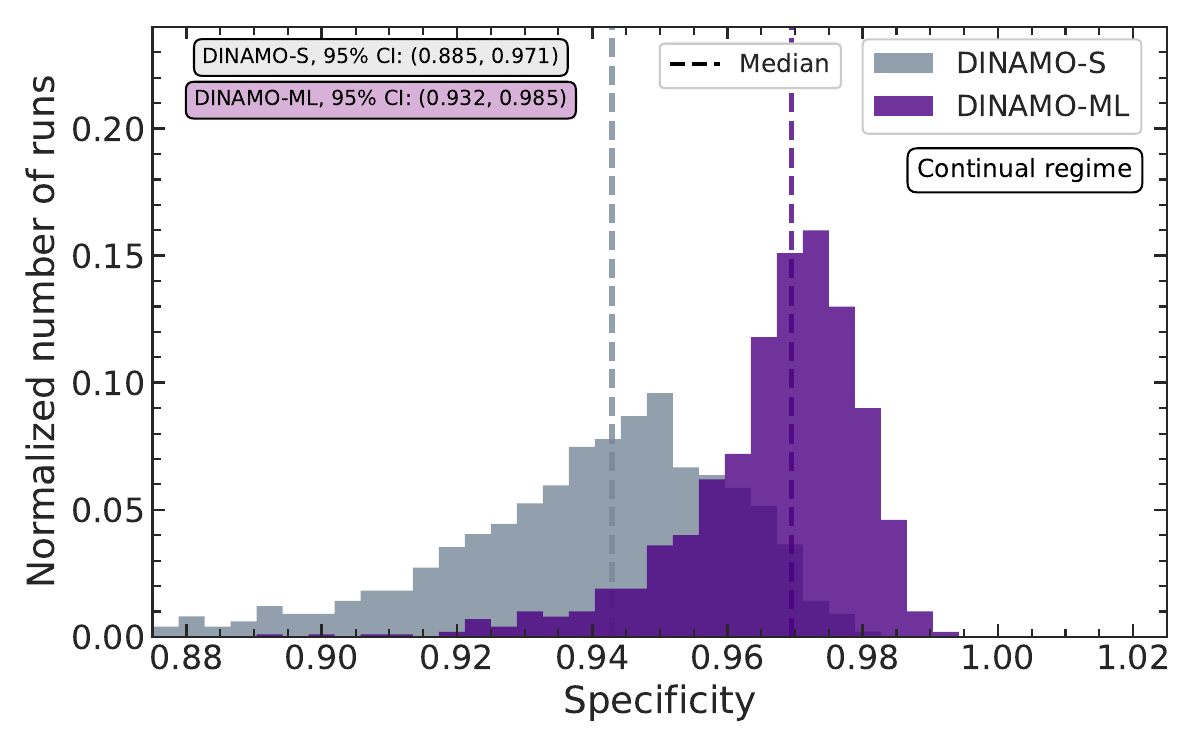}
\caption{Distributions of balanced sensitivity (left) and specificity (right) computed over 1000 different synthetic datasets for both DINAMO-S (in grey) and DINAMO-ML (in violet) methods. The dashed lines represent the medians of the distributions. The metrics are computed in the continual regime only.}
\label{fig:models_comparison_sensitivity_specificity}
\end{figure*}

While DINAMO-ML outperforms DINAMO-S in both sensitivity and specificity metrics, the improvement is particularly notable in specificity as shown in~\autoref{fig:models_comparison_sensitivity_specificity}. This indicates that while both models are effective at detecting anomalies (high sensitivity), DINAMO-S more frequently misclassifies good runs as bad (lower specificity), potentially leading to unnecessary human intervention in a real-world deployment.

\begin{figure}[!hbt]
\begin{center}
\includegraphics[width=0.6\textwidth]{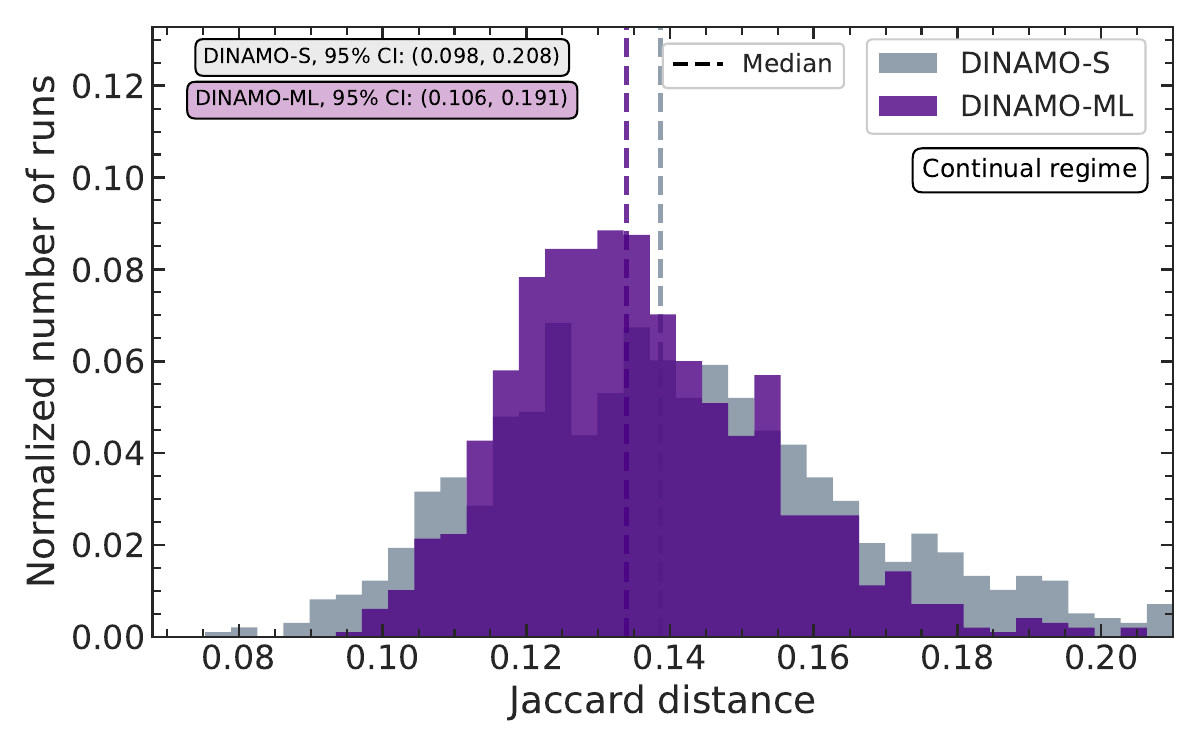}
\caption{Distribution of Jaccard distances computed over 1000 different synthetic datasets for both DINAMO-S (in grey) and DINAMO-ML (in violet) methods. The dashed lines represent the medians of the distributions. The metrics are computed in the continual regime only.}
\label{fig:models_comparison_jd}
\end{center}
\end{figure}

Regarding the models' ability to capture data uncertainty,~\autoref{fig:models_comparison_jd} presents the distribution of the Jaccard distance metric across all 1000 datasets. While the median Jaccard distance values for both models are similar (with DINAMO-ML performing slightly better), the confidence interval for DINAMO-ML is narrower. This indicates that the machine learning-based method provides more consistent uncertainty estimates across different datasets, which could be advantageous in environments with varying operational conditions.

All metrics are summarized in~\autoref{table:agg_results_continual} as median values along with their 95\% confidence intervals for both models. The DINAMO-ML approach outperforms the statistical version, DINAMO-S, across all considered metrics. Its main advantages lie in higher precision and greater adaptability. Note that several ensembling techniques combining anomaly scores from both models were tested but offered no significant improvement over DINAMO-ML alone.

\begin{table*}[t]
\renewcommand*{\arraystretch}{1.25}
\begin{center}
\begin{small}
\begin{sc}
\begin{tabular}{lccccc}
\toprule
 & Bal. acc. $\uparrow$ & Specif. $\uparrow$ & Sensit. $\uparrow$ & Jaccard D. $\downarrow$ & Ad. time $\downarrow$ \\
\midrule
DINAMO-S & $0.947^{+0.020}_{-0.033}$ & $0.943_{-0.058}^{+0.028}$ & $0.956_{-0.075}^{+0.029}$ & $0.139_{-0.041}^{+0.069}$ & $2.02_{-1.13}^{+3.24}$ \\
DINAMO-ML & $\mathbf{0.966_{-0.018}^{+0.012}}$ & $\mathbf{0.969_{-0.037}^{+0.015}}$ & $\mathbf{0.966_{-0.044}^{+0.024}}$ & $\mathbf{0.134_{-0.028}^{+0.057}}$ & $\mathbf{1.61_{-0.61}^{+0.87}}$ \\
\bottomrule
\end{tabular}
\end{sc}
\end{small}
\end{center}
\caption{Proposed quality metrics for the DINAMO-S and DINAMO-ML algorithms in the \textit{continual regime}. The results are aggregated over 1000 different synthetic datasets and reported as medians with the 95\% confidence intervals. \textbf{Bold} denotes the best metric value achieved between the two models.}
\label{table:agg_results_continual}
\end{table*}

\subsection{Stability of the DINAMO-S results with varying smoothing parameter}

\begin{figure*}[!hbt]
\begin{center}
\centerline{\includegraphics[width=\textwidth]{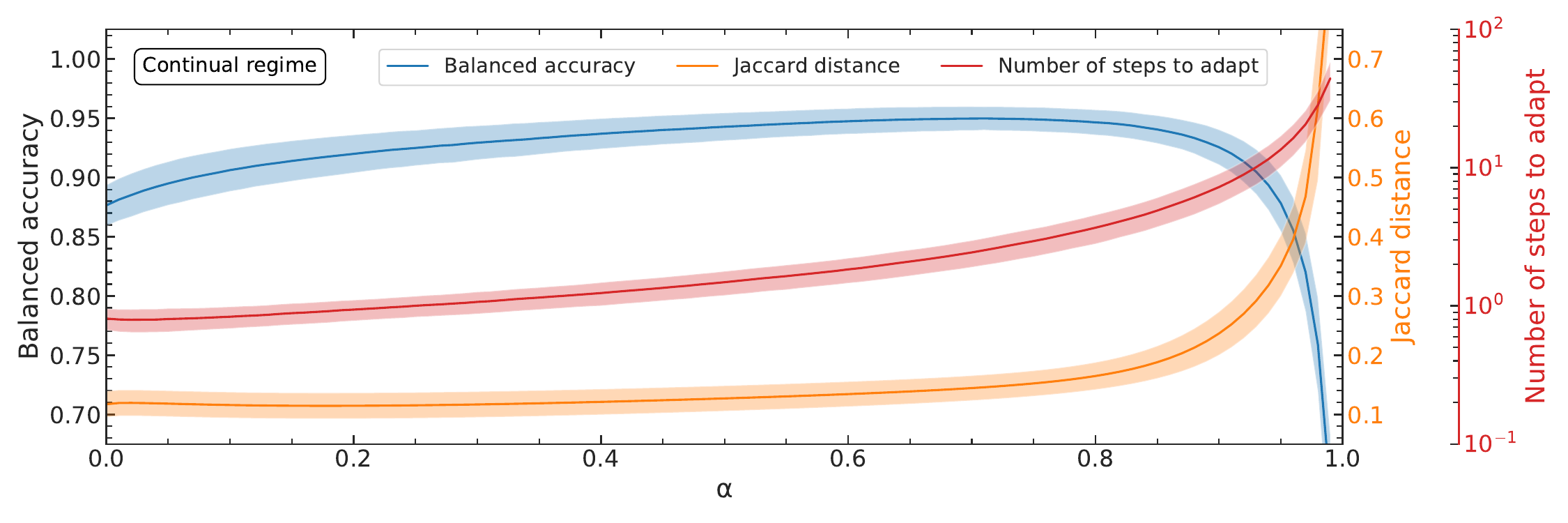}}
\caption{Variation of the performance metrics for the DINAMO-S algorithm as a function of the value of the $\alpha$ parameter. Solid lines (bands) correspond to averages (standard deviations) over the ensemble of 1000 datasets. The performance curves are based on discrete $\alpha$ values from 0.00 to 0.99 with a step size of 0.01 (100 values total).}
\label{fig:metrics_vs_alpha}
\end{center}
\end{figure*}

As introduced in~\autoref{subsec:DINAMO_s}, DINAMO-S is controlled by a single parameter: $\alpha$. This parameter acts as a weight assigned to the ``history'' during the reference update process, while $1 - \alpha$ determines the weight of the most recent good run. Giving more weight to the history allows to reduce the impact of statistical uncertainties in the template creation, while shifting the weight to the most recent histogram allows a quicker adaptation. Hence, one needs to strike a balance between both effects for best accuracy.~\autoref{fig:metrics_vs_alpha} shows how performance metrics vary as a function of $\alpha$, averaged over our ensemble of 1000 datasets.

The performance curves are remarkably flat across a wide range of $\alpha$ values (approximately 0.6 to 0.9), demonstrating the method's robustness even when $\alpha$ is not perfectly optimized for a given dataset. This is particularly important in practical applications, where optimal parameter tuning may not always be feasible.

\subsection{Computing Time.}
The results discussed above were obtained using a machine with 60 Intel(R) Xeon(R) Gold 6326 CPU @ 2.90GHz. For the 1000 datasets, the entire procedure of data generation, visualizations, algorithms' training and results validation took approx. 19 hours parallelized on the 60 CPU cores, where the most time consuming part is the training of the DINAMO-ML model: about 1s for each reference update on a CPU, whereas DINAMO-S requires about 1 ms.

\section{Conclusions}
\label{sec:summary}

We introduced DINAMO, a dynamic and interpretable framework for anomaly detection in Data Quality Monitoring systems at large-scale particle physics experiments. The framework specifically addresses the challenging scenario of frequently changing operational conditions --- a common situation during detector commissioning or following hardware or software modifications. By emphasizing statistical robustness and interpretability, DINAMO offers a practical solution to a long-standing challenge in experimental physics.

The core novelty of DINAMO lies in the construction of evolving histogram references with explicit per-bin uncertainties. This approach offers several distinct advantages:

\begin{itemize}
    \item \textbf{Reduced human workload:} By automatically processing data consistent with references and flagging only anomalous runs, DINAMO dramatically reduces the burden on human shifters.
    
    \item \textbf{Transparent decision-making:} The explicit visualization of references and uncertainties provides a clear basis for validating flagged anomalies, addressing the "black box" problem common in machine learning systems.
    
    \item \textbf{Anomaly localization:} The design naturally extends to multi-histogram inputs, where an overall anomaly score can be decomposed into contributions from individual histograms or detector subsystems, facilitating efficient root-cause analysis.
    
    \item \textbf{Adaptive reference modeling:} The system's ability to update references as conditions change eliminates the need for manual reference building, that can significantly help to reduce time spent by the human experts.
\end{itemize}

The DINAMO framework consists of two complementary methods within. DINAMO-S leverages a novel generalization of Exponentially Weighted Moving Average (EWMA), offering computational efficiency and straightforward implementation. Meanwhile, DINAMO-ML uses transformer encoder-based architecture to model complex temporal dependencies, achieving higher classification accuracy but at the cost of increased computational requirements. Our evaluations on synthetic datasets demonstrate both DINAMO-S and DINAMO-ML yield high balanced accuracies, robust coverage of the reference uncertainties, and short adaptation times (within two iterations on average). The ML-based approach outperforms DINAMO-S across all measured metrics, owing to its ability to more flexibly model temporal dependencies. Although the ML approach requires roughly $\mathcal{O}(10^3)$ more time per iteration than the statistical one, the computational overhead of both algorithms remains negligible compared to the typical multi-minute intervals between runs in particle physics experiments. Consequently, either method can be deployed in real-time DQM.

A key real-world application case is the ongoing commissioning of DINAMO-S by the LHCb experiment for offline DQM in an upcoming data-collection period, following its deployment last year. Each run in LHCb aggregates a set of histograms relevant to monitoring the performance of the various detector subsystems and the software trigger, capturing data features critical for assessing both the detector's performance and reconstruction algorithms over intervals ranging from a few minutes to roughly one hour. DINAMO-S prompts the shifter or subsystem experts with runs that deviate from evolving reference distributions, thus allowing them to narrow their focus to a smaller, potentially problematic subset of data. When an anomaly is detected, the algorithm automatically highlights the ``worst'' histograms (i.e., those with the highest anomaly scores) and presents them alongside their reference templates. By aggregating anomaly scores at the subsystem level, shifters can have a prompt first assessment of what is the source of the deviation. Future publications will explore the full details of the DINAMO-S integration into the LHCb pipeline, including deployment constraints and performance metrics.

Although we have primarily concentrated on a particle physics application, the principles developed here also apply to other domains, such as industrial process control or network security, which experience gradual or abrupt shifts in operational conditions. The adaptability of both DINAMO-S and DINAMO-ML, together with their bin-wise uncertainty coverage, makes them particularly suitable for tasks requiring automatic monitoring with some level of human oversight.

\paragraph{Limitations.} Our reference-construction process models intrinsic uncertainties as Gaussian and statistical uncertainties as Poisson. This pragmatic approximation may require modification if the true data distributions deviate significantly from these forms. Another potential limitation is the frequency at which the reference updates: by updating too often, anomalies that manifest as slow drifts risk being absorbed into the template over time. A practical solution is to cache reference states and recall them at multiple intervals when evaluating each run (e.g., the most recent reference, one from a week prior, one from a month prior). This approach enables simultaneous comparisons across different historical benchmarks, helping to reveal such anomalies. As an additional consideration, the presented algorithms assume binned input data. Extending the framework to unbinned data is a subject for future work. Furthermore, we restricted our study to single-dimensional histograms, as this is standard in many DQM workflows. However, extending to multi-dimensional data is conceptually straightforward by unrolling multi-dimensional bins into one-dimensional representations.

Looking ahead, DINAMO could be extended to the online regime, other particle physics experiments, and broader mission-critical real-time anomaly detection contexts where nominal conditions evolve over time and human oversight and interpretability remain paramount.

\ack
We would like to thank the members of the LHCb Collaboration for useful discussions that helped improve this work. We are also grateful to CloudVeneto for providing IT support and GPU resources. Arsenii Gavrikov is supported by the European Union's Horizon 2020 research and innovation programme under the Marie Skłodowska-Curie Grant Agreement No. 101034319 and from the European Union – NextGenerationEU. Julián García Pardiñas is supported by the U.S. National Science Foundation under Grant No. 2411204. This work has also been supported by the National Resilience and Recovery Plan (PNRR) through the National Center for HPC, Big Data and Quantum Computing.

\section*{References}
%\bibliographystyle{unsrt} %{iopart-num}
%\bibliography{bibliography} % if your bibtex file is called example.bib

\appendix

\clearpage
\section{Synthetic data generation framework}
\label{app:synthetic_data_desc}

To enable systematic evaluation of the DINAMO algorithms, we developed a synthetic data generation framework that captures key characteristics of real-world particle physics data while providing ground truth labels for validation. This appendix describes the technical implementation of this framework, providing details that complement the conceptual overview in the main text.

\paragraph{Overview of the generation process.} We simulate \(N_{\rm runs}\) total runs, each generating a histogram \(\bm{x}_i \in \mathbb{R}^{N_b}\), where $N_b$ is a number of bins. A portion of these runs (\(N_{\mathrm{runs}}^{\mathrm{anomalous}}\)) are labeled bad (\(y_i = 1\)), while the rest are labeled good (\(y_i = 0\)). The label distribution, as well as the time evolution of the underlying distributions, are controlled via several tunable parameters described below.

\paragraph{Slow Drifts.}
We allow gradual drifts in the Gaussian mean \(\mu_i\) across consecutive runs to reflect long-term hardware or environment changes. By default, we define:
\[
\mu_i \;=\;\mu_{\rm{base}} \;+\; A_{\rm{drift}} \,\sin{\Bigl(\frac{i}{\mu_{T_{\mathrm{slow}}}}\,\pi\Bigr)},
\]
where \(\mu_{\rm{base}}\) is an initial mean value and \(A_{\rm{drift}}\) controls the magnitude of sinusoidal fluctuations. The parameter \(\mu_{T_{\mathrm{slow}}}\) sets the timescale (in runs) for one full sine wave.

\paragraph{Rapid Changes.}
With probability \(p_{\mathrm{rapid}}\), we introduce abrupt jumps in \(\mu_i\) or \(\sigma_i\). For example, if a rapid change event is triggered at run \(k\),
\[
\mu_k \;\leftarrow\; \mu_k + \Delta_{\mu},
\quad
\sigma_k \;\leftarrow\; \sigma_k + \Delta_{\sigma},
\]
where \(\Delta_{\mu}\) and \(\Delta_{\sigma}\) are drawn from uniform distributions within user-defined limits, i.e., \([\mu_{\mathrm{min}}^{\mathrm{rapid\_shift}}, \mu_{\mathrm{max}}^{\mathrm{rapid\_shift}})\), \([\sigma_{\mathrm{min}}^{\mathrm{rapid\_shift}}, \sigma_{\mathrm{max}}^{\mathrm{rapid\_shift}})\). These shifts persist until the next rapid change triggers. Such behavior mimics discrete reconfigurations of detector subsystems or hardware resets.

\paragraph{Event Statistics and Binomial Fluctuations.}
Each run \(i\) is assigned an event count \(I_{x,i}\) drawn uniformly from \([I_{\mathrm{min}}, I_{\mathrm{max}}]\). We sample \(\bm{x}_i\) by drawing \(I_{x,i}\) events from \(\mathcal{N}(\mu_i,\, \sigma_i)\) and binning them into \(N_b\) equally spaced bins. To introduce systematic distortions, we optionally apply binomial-like uncertainty  to half of the bins, with binomial probability parameter \(p_{\mathrm{binom}}\).
Specifically, for every run, the right half of the histogram sums with a sample from the binomial distribution, where $n$ is the corresponding bin content. The sum might have either ``+'' sign or ``-'' sign to model either accidental increase of events or accidental decrease of events, mimicking correlated inefficiencies in readout electronics of the detector.

\paragraph{Anomalous Runs.}
We designate \(N_{\mathrm{runs}}^{\mathrm{anomalous}}\) out of \(N_{\mathrm{runs}}\) to be bad. These runs are assigned additional modifications on top of the normal drift and rapid-change mechanics with some probability (\(p_{\mathrm{anomaly}}\), being usually close to 1, so to not have many noisy labels):
\begin{itemize}
\item Extra \(\pm\) shifts in \(\mu_i\) or \(\sigma_i\), drawn from a narrower range than typical rapid changes, where limits are defined in \([\mu_{\mathrm{min}}^{\mathrm{anomaly\_shift}}, \mu_{\mathrm{max}}^{\mathrm{anomaly\_shift}})\), \([\sigma_{\mathrm{min}}^{\mathrm{anomaly\_shift}}, \sigma_{\mathrm{max}}^{\mathrm{anomaly\_shift}})\).
\item Potential ``dead bins,'' where some fraction of bins is forced to zero content, but broader than the slow drifts.
\end{itemize}
We label such runs with \(y_i = 1\); all others remain \(y_i = 0\).

\subsection{Multiple Synthetic Dataset used in the Study}

In this study, we used 1000 different datasets generated by varying random seeds in the stochastic parts of the data generator algorithm. The following parameters of the generation were used: 

\begin{itemize}
    \setlength{\itemsep}{5pt} % Reduce space between items
    \item $N_{\rm runs}$: 5000
    \item $N_{\mathrm{runs}}^{\mathrm{anomalous}}$: 500
    \item $\mu_{\rm{base}}$: 0
    \item $A_{\rm{drift}}$: 0.5
    \item $I_{\mathrm{min}}$: 2000
    \item $I_{\mathrm{max}}$: 20000
    \item $\mu_{T_{\mathrm{slow}}}$: 500
    \item $[\mu_{\mathrm{min}}^{\mathrm{rapid\_shift}}, \mu_{\mathrm{max}}^{\mathrm{rapid\_shift}})$: [0.5, 1.5)
    \item $[\sigma_{\mathrm{min}}^{\mathrm{rapid\_shift}}, \sigma_{\mathrm{max}}^{\mathrm{rapid\_shift}})$: [0.1, 0.4)
    \item $[\mu_{\mathrm{min}}^{\mathrm{anomaly\_shift}}, \mu_{\mathrm{max}}^{\mathrm{anomaly\_shift}})$: [0.25, 0.75)
    \item $[\sigma_{\mathrm{min}}^{\mathrm{anomaly\_shift}}, \sigma_{\mathrm{max}}^{\mathrm{anomaly\_shift}})$: [0.05, 0.2)
    \item $p_{\mathrm{rapid}}$: 0.005
    \item $p_{\mathrm{binom}}$: 0.4
    \item $p_{\mathrm{anomaly}}$: 0.9
\end{itemize}

\autoref{alg:synthetic_data_generator} provides the full pseudocode of the generator. 

\begin{algorithm}[!h]
\caption{Synthetic data generator}
\label{alg:synthetic_data_generator}
\begin{algorithmic}
\STATE {\bfseries Input:} Parameters for data generation
\STATE {\bfseries Initialize:} Empty arrays for histograms and labels
\STATE $\mu_{\rm{current}} \leftarrow \mu_{\rm{base}}$; $\sigma_{\rm{current}} \leftarrow \sigma_{\rm{base}}$
\FOR{$i \leftarrow 1$ {\bfseries to} $N_{\rm runs}$}
    \STATE $\mu_{\rm{slow\_drift}} \leftarrow \mu_{\rm{base}} + A_{\rm{drift}} \cdot \sin(\pi \cdot i / \mu_{T_{\mathrm{slow}}})$ {\bfseries // Apply slow drift to $\mu$}
    \IF{$\rm{random}() < p_{\mathrm{rapid}}$}
        \STATE $\rm{sgn_{\mu}}$, $\rm{sgn_{\sigma}} \leftarrow \rm{random\_choice}(\{-1, 1\})$
        \STATE $\mu_{\rm{current}} \leftarrow \mu_{\rm{slow\_drift}} + \rm{sgn_{\mu}} \cdot \rm{uniform}(\mu_{\mathrm{min}}^{\mathrm{rapid\_shift}}, \mu_{\mathrm{max}}^{\mathrm{rapid\_shift}})$
        \STATE $\sigma_{\rm{current}} \leftarrow \sigma_{\rm{current}} + \rm{sgn_{\sigma}} \cdot \rm{uniform}(\sigma_{\mathrm{min}}^{\mathrm{rapid\_shift}}, \sigma_{\mathrm{max}}^{\mathrm{rapid\_shift}})$
    \ELSE
        \STATE $\mu_{\rm{current}} \leftarrow \mu_{\rm{slow\_drift}}$
    \ENDIF    
    \STATE $\rm{is\_anomalous} \leftarrow i \in$ Randomly selected indices with $p_{\mathrm{anomaly}}$
    \STATE $y_i \leftarrow 0$  {\bfseries // Default to good run}
    
    \IF{$\rm{is\_anomalous}$} 
        \STATE $\rm{sgn_{\mu}}$, $\rm{sgn_{\sigma}} \leftarrow \rm{random\_choice}(\{-1, 1\})$

        \STATE $\mu_{\rm{anomaly}} \leftarrow \mu_{\rm{current}} + \rm{sgn_{\mu}} \cdot \rm{uniform}(\mu_{\mathrm{min}}^{\mathrm{anomaly\_shift}}, \mu_{\mathrm{max}}^{\mathrm{anomaly\_shift}})$
        \STATE $\sigma_{\rm{anomaly}} \leftarrow \sigma_{\rm{current}} + \rm{sgn_{\sigma}} \cdot \rm{uniform}(\sigma_{\mathrm{min}}^{\mathrm{anomaly\_shift}}, \sigma_{\mathrm{max}}^{\mathrm{anomaly\_shift}})$
      
        \STATE $\rm{dead\_bins} \leftarrow \rm{randomly\_select\_bins}(N_b, \rm{fraction\_dead\_bins})$
        \STATE $y_i \leftarrow 1$  {\bfseries // Mark as bad run}
    \ELSE
        \STATE $\mu_{\rm{anomaly}} \leftarrow \mu_{\rm{current}}$; $\sigma_{\rm{anomaly}} \leftarrow \sigma_{\rm{current}}$
        \STATE $\rm{dead\_bins} \leftarrow \{\}$
    \ENDIF
    
    \STATE $I_{x_i} \leftarrow \rm{uniform}(I_{\mathrm{min}}, I_{\mathrm{max}})$
    \STATE $\rm{events} \leftarrow \rm{sample\_from\_gaussian}(\mu_{\rm{anomaly}}, \sigma_{\rm{anomaly}}, I_{x_i})$
    \STATE $\bm{x}_i \leftarrow \rm{bin\_events}(\rm{events}, N_b, \rm{bin\_range})$
    
    \FOR{$j \leftarrow N_b/2$ {\bfseries to} $N_b$}
        \IF{$\bm{x}_i[j] > 0$}
            \STATE $\rm{sgn_{binom}} \leftarrow \rm{random\_choice}(\{-1, 1\})$
            \STATE $\bm{x}_i[j] \leftarrow \bm{x}_i[j] + \rm{sgn_{binom}} \cdot \rm{sample\_binomial}(\bm{x}_i[j], p_{\mathrm{binom}})$
        \ENDIF
    \ENDFOR
    
    \FOR{$j \in \rm{dead\_bins}$}
        \STATE $\bm{x}_i[j] \leftarrow 0$ {\bfseries // Apply dead bins}
    \ENDFOR
    
    \STATE $\rm{histograms}[i] \leftarrow \bm{x}_i$; $\rm{labels}[i] \leftarrow y_i$
\ENDFOR

\STATE {\bfseries Output:} Histograms $\{\bm{x}_i\}_{i=1}^{N_{\rm runs}}$ and labels $\{y_i\}_{i=1}^{N_{\rm runs}}$
\end{algorithmic}
\end{algorithm}

\clearpage
\section{DINAMO-S algorithm details}
\label{app:details_dinamo_s}

\autoref{alg:dinamo_s} of this appendix presents the full preudocode of the DINAMO-S algorithm.

\begin{algorithm}[h]
   \caption{DINAMO-S: EWMA-based algorithm}
   \label{alg:dinamo_s}
\begin{algorithmic}
   \STATE {\bfseries Input:} Histograms $\{x_{i}\}_{i=1}^N$, labels $\{y_{i}\}_{i=1}^N$, EWMA factor $\alpha$, small $\varepsilon>0$
   \STATE
   \STATE {\bfseries Initialize:}
   \STATE \quad $(\bm{\mu}_0, \bm{\sigma}_{\mu_0, p}, I_{\mu_0}) \leftarrow \textsc{ReferenceInit}()$ \quad(\textit{uniform histogram bins, then normalized})
   \STATE \quad $(\bm{W}_0, \bm{S}_{\mu_0}, \bm{S}_{\sigma_\mu, 0}) \leftarrow \textsc{EWMAInit}(\alpha, \mu_0, \sigma_{p,\mu_0})$
   \STATE
   \FOR{$i \leftarrow 1$ {\bfseries to} $N$}
       \STATE $I_{x_i} \leftarrow \sum_{j=1}^{N_b} x_{i, j}$
       \IF{$I_{x_i} > \varepsilon$}
          \STATE $\bm{\tilde{x}}_{i} \leftarrow \bm{x}_{i} / I_{x_i}$ \quad(\textit{normalize to unity})
          \STATE $\bm{\sigma}_{\bm{\tilde{x}}_i, p} \leftarrow \textsc{PoissonUncertainty}(\bm{\tilde{x}}_i)$
       \ENDIF
       \STATE
       \STATE \textbf{Compute} \(\chi^2\) and pull $\bm{\delta}$
       \STATE
       \STATE \textbf{If} $y_i = 0$ \textbf{ (good run) then}:
       \STATE \quad $\bigl(\bm{\mu}_{i+1}, \bm{\sigma}_{\mu, {i+1}}, \bm{W}_{i+1}, \bm{S}_{\mu, i+1}, \bm{S}_{\sigma_\mu, i+1}\bigr) \,\leftarrow\, \textsc{UpdateReference}\bigl(\alpha, \bm{\tilde{x}}_{i}, I_{x_i}, \bm{\sigma}_{\bm{\tilde{x}}_i, p}, \bm{\mu}_i, \bm{\sigma}_{\mu, i}, \bm{W}_i, \bm{S}_{\mu, i}, \bm{S}_{\sigma_{\mu}, i}\bigr)$
       \STATE
       \STATE \textbf{Store} \(\chi^2\), pull $\bm{\delta}$, updated \(\bm{\mu}_{i+1}\) and \(\bm{\sigma}_{\mu, i+1}\)
   \ENDFOR
   \STATE
   \STATE {\bfseries Output:} Lists of references with their uncertainties, per-run \(\chi^2\) and pull $\bm{\delta}$ values
\end{algorithmic}
\end{algorithm}

\clearpage
\section{DINAMO-ML algorithm details}
\label{app:details_dinamo_ml}

\begin{figure*}[!htb]
\begin{center}
\includegraphics[width=1\textwidth]{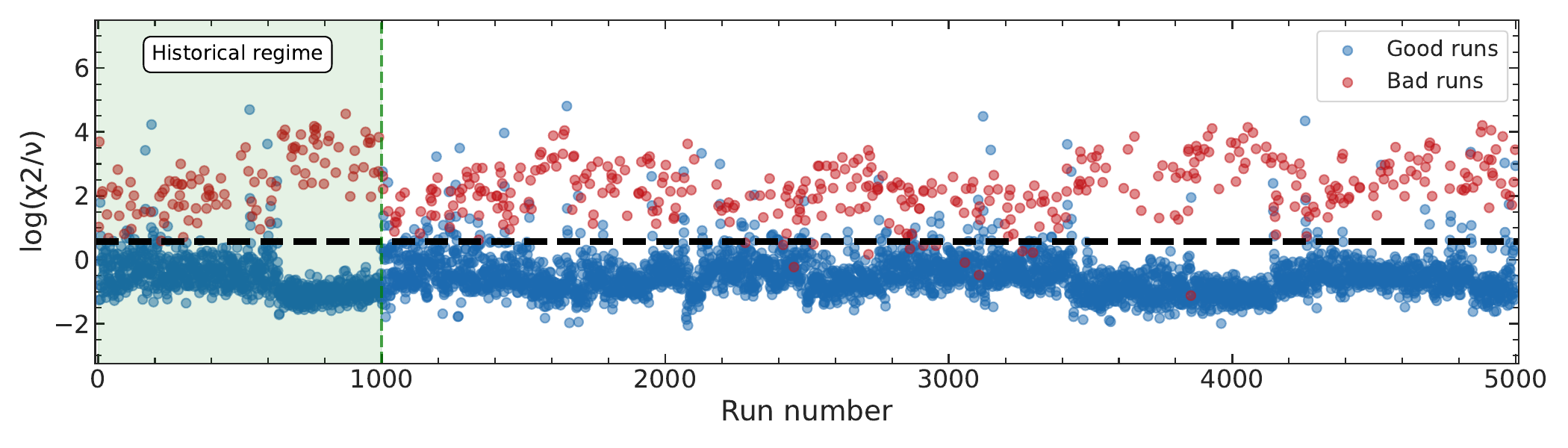}
\caption{Anomaly score of the DINAMO-ML algorithms as a function of the run number for a single dataset composed of 5000 runs. Red points represent the ground-truth bad runs and the blue points represent the good runs. The black dashed line is the optimized, using the historical regime, threshold to assign a class label: if a run is above it, the run is predicted to be bad and vice versa.}
\label{fig:dinamo_ml_scatter_plot}
\end{center}
\end{figure*}

This appendix presents some additional information related to the machine learning-based DINAMO model.
~\autoref{fig:dinamo_ml_scatter_plot} shows the logarithm of the reduced $\chi^2$ anomaly score for each run. Similarly to the DINAMO-S presented on~\autoref{fig:dinamo_s_scatter_plot}, DINAMO-ML has proven to be highly accurate in detecting bad runs as well as to quickly adapt to changing conditions.  

\autoref{alg:dinamo_ml} shows the full pseudocode of the DINAMO-ML method. It details the initialization process, the context-building procedure, the inference steps for anomaly detection, and the training mechanism that allows the model to adapt over time as new good runs are observed. 

\begin{algorithm}[h]
\caption{DINAMO-ML: Transformer encoder-based algorithm}
\label{alg:dinamo_ml}
\begin{algorithmic}
  \STATE {\bfseries Input:} $\{\bm{x}_{i},y_{i}\}_{i=1}^N$, buffer size $M$, batch size $K$, learning rate, etc.
  \STATE {\bfseries Initialize:}
  \STATE \quad $\mathcal{D} \leftarrow \varnothing$
  \STATE \quad $\rm{NNParams} \leftarrow \textsc{InitTransformerEncoder}()$
  \STATE \quad $\rm{Optimizer} \leftarrow \textsc{AdamW}(\rm{NNParams}, \rm{lr}=5\times10^{-4}, \rm{weight\_decay}=10^{-4})$
  \STATE \quad \textsc{EarlyStopPatience} $\leftarrow 5$

  \FOR{$i \leftarrow 1$ {\bfseries to} $N$}
    \STATE \textbf{// 1. Normalize the new histogram}
    \STATE Normalize $\bm{x}_i$ (if nonzero), get $\bm{\tilde{x}}_i$

    \STATE \textbf{// 2. Perform inference to obtain reference prediction}
    \STATE $\rm{context} \leftarrow \textsc{BuildContext}(\mathcal{D},\, i,\, M)$
    \STATE $(\hat{\mu}_i, \hat{\sigma}_i) \leftarrow \textsc{TransformerEncoderForward}(\rm{context},\, \rm{NNParams})$
    \STATE \textbf{// 3. Compute $\chi^2_\nu$ and compare to anomaly threshold to classify}

    \STATE \textbf{// 4. A posteriori label is assigned (by shifter or ground truth)}
    \IF{$y^{(i)} = 0$} 
      \STATE $\mathcal{D} \leftarrow \mathcal{D} \cup (\bm{\tilde{x}}_i, i)$

        \STATE \textbf{// 5. Training step}
        \STATE \textbf{Select} the $K$ most recent entries from $\mathcal{D}$ (if fewer than $K$ exist, take all).
        \STATE $nll_{\rm{batch}} \leftarrow 0$
        \STATE \textsc{Optimizer.zero\_grad}()
    
          \STATE \textbf{for each} $(\bm{\tilde{x}}_{k}, i_k)$ \textbf{in the selected batch}:
            \STATE \quad $\rm{context} \leftarrow \textsc{BuildContext}(\mathcal{D},\, i_k,\, M)$
            \STATE \quad $(\hat{\bm{\mu}}, \hat{\bm{\sigma}}) \leftarrow \textsc{TransformerEncoderForward}(\rm{context},\, \rm{NNParams})$
            \STATE \quad $nll_{\mathrm{k}} \leftarrow \textsc{ComputeNLL}(\bm{\tilde{x}}_k, \hat{\bm{\mu}}, \hat{\bm{\sigma}})$
            \STATE \quad $nll_{\rm{batch}} \leftarrow nll_{\rm{batch}} + nll_{\mathrm{k}}$
    
          \STATE $nll_{\rm{batch}} \leftarrow nll_{\rm{batch}} \;\big/\; \#(\rm{selected batch})$
          \STATE \textsc{Backprop} on $nll_{\rm{batch}}$
          \STATE \textsc{CheckEarlyStopping}(\textsc{EarlyStopPatience})
    \ENDIF

    \STATE \textbf{/* End of iteration for run $i$ */}
  \ENDFOR
\end{algorithmic}
\end{algorithm}

\end{document}